\documentstyle[12pt,aaspp4]{article}

\begin{document}

\title{Models for Dense Molecular Cloud Cores}

\author{Steven D. Doty and David A. Neufeld} 
\affil{Department of Physics and Astronomy, The Johns Hopkins University\\
3400 N.~Charles St., Baltimore, MD  21218}

\begin{abstract}
We present a detailed theoretical model for the thermal 
balance, chemistry, and radiative transfer within quiescent
dense molecular cloud cores that contain a central
protostar.  In the interior of such cores, we expect the 
dust and gas temperatures to be well coupled, while in
the outer regions CO rotational emissions dominate the gas 
cooling and the predicted gas temperature lies
significantly below the dust temperature. 
Large spatial variations in the gas temperature are expected
to affect the gas phase chemistry dramatically; in particular, the
predicted water abundance varies by more than a factor of 1000 
within cloud cores that contain luminous protostars. 

Based upon our predictions for the thermal and chemical
structure of cloud cores, we have constructed self-consistent
radiative transfer models to compute the line strengths 
and line profiles for transitions of $^{12}$CO, $^{13}$CO, C$^{18}$O,
ortho- and para-H$_{2}^{16}$O, ortho- and para-H$_2^{18}$O,
and OI.  We carried out a general parameter study to
determine the dependence of the model predictions upon the
parameters assumed for the source. We expect many of the 
far-infrared and submillimeter rotational
transitions of water to be detectable either in emission or 
absorption with the use of the {\it Infrared Space Observatory}
(ISO) and the {\it Submillimeter Wave Astronomy Satellite}.
Quiescent, radiatively-heated hot cores are expected to show
low-gain maser emission in the 183 GHz $3_{13}-2_{20}$ water
line, such as has been observed towards several hot core 
regions using ground base telescopes.
We predict the $\rm ^3P_1 - ^3P_2$ fine structure transition
of atomic oxygen near 63 $\mu$m to be in strong absorption
against the continuum for many sources.
Our model can also account successfully for recent ISO observations
of absorption in rovibrational transitions of water toward
the source AFGL 2591.  

\end{abstract}


\section{Introduction}

The role of dense molecular clouds as the sites of star formation has been
clearly established (e.g.\ Shu, Adams, \& Lizano 1987, and references
therein).
Although the evolution of cloud cores from the onset of
collapse to fragmentation and star formation is not completely understood,
the collapse process is partly regulated by the ability of the
core to cool itself.  A full understanding of the thermal balance
within molecular cloud cores requires a careful consideration of
the microphysics of molecular excitation, cloud core chemistry,
and radiative transfer.

Goldsmith and Langer (1978) carried out a comprehensive study of
radiative cooling in dense molecular clouds, computing the cooling rates
from a number of molecules as a function of molecular hydrogen density and
gas temperature. Neufeld, Lepp, and Melnick (1995, hereafter NLM)\
re-examined the cooling rates using the molecular cooling functions obtained by
Neufeld and Kaufman (1993, hereafter NK) and included the effects of
gas phase chemistry. They also predicted line
strengths for many important species (e.g. H$_2$O \& CO)\ 
for the ideal case of an isothermal cloud core in hydrostatic
equilibrium, and argued that much of the energy radiated by the
gas in molecular cloud cores emerges in far-infrared and
submillimeter transitions that are detectable only from satellite
observatories.   More recently,  Ceccarelli, Hollenbach \& Tielens
(1996, hereafter CHT) 
have presented a model for collapsing cloud cores 
in which the chemistry, thermal balance, 
and line emission were treated 
simultaneously.

These recent theoretical developments have been motivated in large
part by rapid advances in observational capabilities.  The recent launch
of the {\it Infrared Space Observatory} and the planned launch
of the {\it Submillimeter Wave Astronomy Satellite} (SWAS) will
allow the far-infrared and submillimeter spectral regions to be probed
without any hindrance by the Earth's atmosphere.  
These satellite observatories will therefore
afford us an ideal opportunity to study transitions
for which the atmospheric absorption is severe, such as
molecular oxygen transitions and far-infrared water transitions.
Although emission in several submillimeter water transitions has 
been observed  from ground-based telescopes or airborne observatories (e.g.,
Waters et al. 1980; Phillips et al. 1980; Wannier et al. 1991;
Jacq et al. 1988; Cernicharo et al. 1990, 1994; Gensheimer et al. 1996;
Zmuidzinas et al. 1996; Tauber et al. 1996), such observations
typically probe transitions which play a negligible role in the
thermal balance of the emitting gas.  SWAS and ISO, by contrast,
will allow the dominant cooling transitions of dense molecular
cloud cores to be studied.
ISO will also allow vibrational bands of molecules to be observed 
in {\it absorption} toward bright infrared sources.
For example, recent ISO 
observations by Helmich et al. (1996), and by van Dishoeck \& Helmich (1996)
of the $\nu_2$ vibrational band of water in absorption
toward several protostars provide an important constraint on 
the abundance of water relative to other species.  

Motivated by these observational developments, we have
constructed a detailed theoretical model for dense molecular
cloud cores, obtaining a self-consistent solution for the 
thermal balance, chemistry, and
radiative transfer at every point within a spherical cloud core. 
After determining the radial distribution of the  dust and gas temperatures and
the chemical composition, we computed self-consistently the
line strengths and line profiles
for relevant transitions of astrophysically important molecules.  We
have considered a large grid of models in order to explore a wide range of
possible source parameters;
in particular, the present study extends the work of CHT to the case of
cloud cores with embedded protostars of very high luminosity.

In \S 2 we describe our model for dense molecular cloud cores 
and discuss the microphysical processes that were included. 
In \S 3, we discuss the results obtained for a standard set of
parameters relevant to a hot core region such as the Orion Molecular
Cloud 1.  In \S 4 we consider a grid of models covering the
entire range of cloud core masses and protostellar luminosities
that are applicable to star-forming regions in the Galaxy. 
In \S 5 we discuss 
the results of this parameter study
in light of current and upcoming observational capabilities.
A brief summary follows in \S 6.

%

\section{Model}
We have modeled the dense, quiescent, star-forming cores of 
molecular clouds.
We assume that the cores are spherical, and
-- as our standard assumption --
that the molecular hydrogen
density
follows an $r^{-2}$ distribution, consistent with both observations
(Goldsmith et al. 1980, Keto et al. 1988, Myers \&\ Fuller 1992, Fuller \&\
Myers 1993) and the fact that theoretical models of star formation (e.g.
Larson 1969) predict evolution toward an $r^{-2}$ distribution even for a
homogeneous initial state.
We have also carried out additional test calculations with an alternate
$r^{-1.5}$ density distribution, with results which are described
in \S3.6 below.
The cloud core is assumed to have inner and outer radii at which the
H$_2$ densities are $10^9$ cm$^{-3}$ and $10^3$ cm$^{-3}$ respectively,
although these limits can be changed significantly with little
effect on the emission for most transitions.
The microturbulent doppler parameter, $\Delta v$, and the exact
density profile are determined from considerations of hydrostatic
equilibrium as discussed by Neufeld \& Green (1994).
Given the standard $r^{-2}$ density distribution, we are left
with two source parameters:  the gas density at some
reference point (or equivalently the total mass
$M(<d)$, enclosed within a region of diameter $d$), and the
internal source luminosity, $L_{*}$.  

Once the mass enclosed (and hence gas density)\ and source luminosity,
have been specified, 
the temperature structure and line emission may be
determined.  Our model -- represented schematically in Figure 1 --
includes careful consideration of the thermal balance, chemistry and
radiative transfer within the cloud core. These are discussed below.

\subsection{Thermal Balance}

The equilibrium gas temperature within the cloud core
is determined by the balance between heating and cooling processes.

The gas heating is dominated by gas-grain collisional heating.
Dust grains within the core are heated both from the inside by 
the central protostar and from the outside by the interstellar
radiation field;  inelastic collisions then
transfer energy from the warmer dust grains to the gas.
We determined the dust temperature from a 
self-consistent solution of the continuum radiative transfer problem
(\S2.3 below), and then used the
estimate of Hollenbach \& McKee (1989) for the rate of 
collisional energy transfer to the gas.

We adopted the cooling functions of NLM to determine the
gas cooling rates due to emissions from CO, O$_2$, OI, H$_2$, 
H$_2$O as well as from other diatomic and polyatomic molecules.
The NLM results are based upon the use of an approximate escape probability
method to treat the effects of radiative trapping in optically
thick cooling transitions, and neglect the effects of pumping
by infrared continuum radiation.
They have been tabulated by
NLM as a function of
the gas density, the gas temperature, and an optical depth parameter
denoted $\widetilde{N}$.

In the outer regions of dense cloud cores, where radiative cooling causes
the gas temperature to fall significantly below the dust temperature,
carbon monoxide is the dominant coolant. We have therefore used a 
full radiative transfer
code employing the method of approximate lambda iteration (ALI, c.f.\
\S 2.4 below) to determine the CO cooling in detail for comparison
with the NLM cooling functions.  We find that exact CO cooling rate is
well approximated by the expression
\begin{equation}
\Lambda _{\rm CO}= \biggl[ \Lambda _{\rm NLM}(T_{\rm gas
})-\Lambda _{\rm NLM}(2.7 K) \biggr]\label{coolminuscbr}
\end{equation}
evaluated for an optical depth parameter (c.f.\ NK)
\begin{equation}
\widetilde{N} = 0.5 N_{\rm CO}/ \Delta v,
\end{equation}
where $\Lambda _{\rm NLM}$ is the NLM cooling function for CO,
$N_{\rm CO}$ is the CO column density along a radial path to
the cloud surface, and $\Delta v$ is the Doppler parameter.
The second term in equation (1) accounts for the effects
of heating by the 2.7~K cosmic background radiation.  Note also that
the best fit to the ALI results was obtained with the optical depth
parameter taken as $0.5 N_{\rm CO} /\Delta v$, a factor of 2
below the value recommended by NK for the case of a spherical
cloud with a $r^{-2}$ density profile.  With these modifications
to the NLM cooling function, we derived an equilibrium gas temperature
that agreed everywhere to within 1~K with the results
obtained using the ALI radiative transfer code to treat CO cooling exactly.

Similar ALI calculations were also carried out for water and OI.
They showed that in cores with an embedded protostar of high luminosity,
{\it absorption} by water and atomic oxygen could actually lead to
{\it net} heating of the gas (confirming a conclusion reached 
earlier by Takahashi, Hollenbach \& Silk (1983) for the case of
water transitions in hot core regions).  However, water and
OI transitions were found to have a significant effect upon the radiative
cooling rate only deep in the cloud interior where the gas temperature
was in any case very closely coupled to the dust temperature.

In the very outer parts of the cloud, grain photoelectric heating
and cosmic ray heating may be significant relative to gas-grain
collisional heating.  Neither of these processes is included in
our present study, so the gas temperature near the outer edge
of the cloud may have been somewhat underestimated.

\subsection{Chemistry}

We adopted the steady-state molecular abundances that were 
computed by NLM, who included the full UMIST chemical network (Millar
et al. 1991, Farquhar \&\ Millar 1993) and assumed the initial gas-phase 
elemental abundances of Millar et al. (1991, Table 4).  
While this
approach has the benefit of being easily calculable, the
consideration of only steady-state gas-phase chemistry 
may underestimate the abundances of
volatile ices such as water and methanol that have been recently 
vaporized from grain surfaces in hot core regions.
On the other hand, water is predicted to be very abundant in
any case for gas temperatures greater than $\sim 300$~K, so the
neglect of time-dependent 
and grain-surface
chemical effects will have a significant
effect upon the predicted water abundance only in a relatively 
small region where the gas temperature is smaller than 300~K
but the dust temperature exceeds the sublimation temperature of
water ice ($\sim 100$~K).
As our standard assumption, we therefore ignore 
the complications of time-dependent chemical
effects and the desorption of icy grain mantles in the present study. 
We have, however, performed test calculations to investigate
what effects upon the line strengths are likely to result
from the desorption of icy grain mantles above 100 K; as
described in \S 3.6 below, the effects are relatively small.

\subsection{Continuum Radiative Transfer}

A detailed treatment of the radiative transfer through dust
is necessary to correctly determine the dust temperature
(and hence the gas temperature; see \S 2.1), as well 
as the infrared pumping and emergent spectrum.  
We adopt dust properties that are based upon Draine's (1987) astronomical
silicate in the visible through mid-infrared spectral
regions ($\lambda <25$ $\mu $m), with an assumed
grain size of $a=0.1$ $\mu $m. For $\lambda >25$ $\mu $m, we assume 
a dust opacity $\propto \lambda ^{-1.5}$ for all
models, and we adopt a normalization for the
dust opacity corresponding to $N({\rm H}_2)/A_{V}\sim 2\times
10^{21}\,\rm cm^{-2}$  (Bohlin et al 1978, Whittet 1992).
Given specified parameters for the cloud core
(viz.\ $L_{*}$, $M_{0.1}$) and these assumed optical properties 
for the dust, we used a modified version of the continuum radiative transfer
code of Egan, Leung, \&\ Spagna (1988) to solve
self-consistently for the
dust emissivity, opacity, and temperature as functions of position.

\subsection{Line Radiative Transfer}

Given the radial distribution of gas density, molecular abundance, and gas
temperature, we are able to determine the level populations and emergent 
line intensities for molecules of interest in hot core regions. This is
accomplished through the self-consistent solution of the radiation transport
problem in 1-D spherical geometry using a code (Doty 1997) 
based on the approximate lambda iteration (ALI)\ method proposed by Rybicki
\&\ Hummer (1992). Along with the spatial variations in the gas parameters,
our code also includes the effects of dust absorption and emission (as
determined from the continuum radiative transfer model).

In formulating the equations of statistical equilibrium for the molecular
level populations, we adopted the same spontaneous radiative rates and
collisional rate coefficients used by NLM, except in the case of
water, for which new estimates of the collisional rate coefficients
have recently become available.
In particular, Phillips et al.\ (1995) have obtained the first estimates
of the inelastic collision cross-sections for the H$_2$--H$_2$O system, 
yielding rate coefficients at temperatures $T<140$ K for collisionally
induced transitions among the lowest ten rotational states of ortho--
and para-water
as a result of collisions with either H$_2$ ($J=0$) or H$_2$ ($J=1$).
Thus whereas NLM simply assumed that the earlier collision cross-sections
computed by Green, Maluendes, and McLean (1993) for the He--H$_2$O
system applied also to H$_2$--H$_2$O, in the present study we have used the
actual results of Phillips et al.\ (1995) for H$_2$--H$_2$O collisions.  
Assuming that cloud cores were initially cold ($T<20$K) and in
local thermodynamic equilibrium, and that the time since the formation 
of an internal heat source (i.e.\ protostar)
has been short in comparison to the para-H$_2$ to
ortho-H$_2$ conversion time ($>10^6$ yr according to 
Abgrall et al. 1992), we expect that molecular hydrogen will
be almost entirely in the para ($J=0$) state.  We therefore adopted the rate coefficients computed by
Phillips et al.\ (1995)
for the excitation of water by H$_2$ ($J=0$); these typically lie
within a factor 2 of the corresponding values for excitation of H$_2$O
by He.  
For comparison, we have also performed test calculations
with an assumed 3:1 ratio of ortho- to para-H$_2$; as described
in \S3.6, the strengths of most water lines show only a 
relatively weak dependence on the assumed ortho- to para-H$_2$ ratio.

In the calculations presented here, we have included just
nine transitions involving the lowest seven rotational states for 
both ortho- and para-water. 
The relatively small number of water states included in our standard model
makes it practical to compute a large grid of models.  Test calculations
involving up to 45 rotational states have shown that our seven-level
model yields predicted line strengths for the nine transitions we
considered that lie well within
5\% of those derived from models that include a much larger number of 
rotational states.

\section{Results for the Standard Model}

We have constructed a computer model to solve simultaneously for
the gas and dust temperatures, molecular abundances and radiative transfer
within a static dense molecular cloud core with an $r^{-2}$ density profile.
In this section (\S 3)
we describe the results obtained for a ``standard" model with
parameters applicable to a high luminosity ``hot core" region.
In the next section (\S 4) we present the results of a general
parameter study.

Our cloud core model has two input parameters:  the mass
enclosed in a given diameter, $M(<d)$, and the internal 
source luminosity, $L_*$.
Here we discuss a ``standard" model with 
$M_{0.1}\equiv M(<0.1$ pc$)=60$ M$_{\odot }$ and 
$L_{*}=10^5$ L$_{\odot }$.
Although this model has parameters
similar to the Orion hot core, direct comparison of our model with
observations of Orion is not possible due to the extremely complicated nature
of the Orion region. For example, our present level of detail cannot include
the effects of shocks, molecular outflows, irregular geometry,
clumping, and multiple heat sources -- all of which are known to 
exist in the Orion region (Genzel \&\ Stutzki 1989 and references 
therein).  We expect however, that our results can be compared to 
observations for both simpler, more regular sources, and for 
some general source-averaged properties.

%

\subsection{Dust temperature distribution}

As noted earlier, we self-consistently solve the radiative
transfer problem through the dust.
In Figure 2, we compare the resulting model continuum spectrum (dashed line),
to observations of Orion (Erickson et al. 1977, and references therein;
filled circles) for the cloud core parameters of the standard model. 
It should be noted that the mid- and far-infrared opacity 
($\propto \lambda ^{-1.5}$) we have assumed
(\S 2.3)
provides a significantly
better fit to the observational data than does the dust opacity 
of Draine (1987). 

In addition to determining the infrared continuum emission, the 
dust temperature affects the gas-grain heating rate (c.f.\
\S 2.1). In Figure 3 we plot the dust temperature (dashed line)\ 
as a function of radial position for this model. The dust in the interior
is heated by the enclosed radiation source, while the dust on the outside is
heated by the interstellar radiation field (ISRF), which we take from
Mathis, Mezger \&\ Panagia (1983). The dip in the dust temperature near 
$r \sim 3-5\times 10^{16}$ cm is due to the fact that at this point 
the thermal dust radiation can begin to escape easily, thereby cooling
the dust and decreasing the temperature.

\subsection{Gas temperature distribution}

Once the source properties and dust temperature distribution have
been specified,  
the gas temperature distribution may be determined by considerations
of thermal balance as described in
\S 2.1.  The resulting gas
temperature profile is plotted in Figure 3 (solid line) for the
standard model. 
The dust and gas temperatures are well coupled in the interior of the 
cloud core (for $r<10^{17}$ cm), because the gas-grain collisional
energy transfer is very effective at high densities and because
the large column densities to the cloud surface lead to
radiative trapping of the molecular line emissions that cool the gas.
At larger radii, line emissions can escape more readily and 
the effectiveness of gas-grain heating diminishes, 
thereby allowing the gas
and dust to decouple thermally and the gas temperature to drop
significantly below the dust temperature.

\subsection{Abundances}

Simultaneously with the gas temperature, the gas phase chemical abundances are
also determined using the 
steady-state
solution of the UMIST 
reaction set (see Figure 1). The resulting fractional abundances for our
standard model are shown in Figure 4, for CO (solid line), OI 
(dotted line), and ortho-H$_2$O (dashed line).  The predicted water 
abundance, in particular, shows a strong spatial variation.
In the cloud interior, where the gas temperature is greater
than $300$ K, water is
formed rapidly by the 
neutral-neutral reactions:

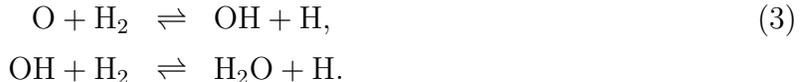
\begin{eqnarray}
{\rm O}+{\rm H}_2 &\rightleftharpoons & {\rm OH}+ {\rm H},
\label{waterformrxn1} \\
{\rm OH}+ {\rm H}_2 &\rightleftharpoons & {\rm H}_{2}{\rm O}+ {\rm H}. 
\nonumber
\end{eqnarray}
This reaction sequence accounts
for the large predicted abundance of water for $r<2-3\times
10^{16}$ cm in our model, and the correspondingly low abundances of atomic
oxygen.
In the cool exterior, however, 
the forward reactions in equation (\ref{waterformrxn1})
are slow due to the substantial activation energies involved, and the water
abundance is significantly smaller;
at low temperatures, water is formed only by a
less effective ion-molecule reaction route
that is driven by cosmic ray ionization.

\subsection{Molecular line cooling}

In Figure 5, we show the cooling coefficients
($\equiv \Lambda /n^2[{\rm H}
_2]$) for the various important coolants in our model as a function of
radial position. CO dominates the cooling throughout most of the cloud
core (i.e.\ for $r>2\times 10^{16}$ cm), with O$_2$, polyatomic molecules, and
other diatomic molecules playing increasingly smaller roles.
For $r<1-2\times 10^{16}$ cm, H$_2$ and polyatomic molecules
dominate the cooling, the high gas temperature allowing the effective
excitation of H$_2$ vibrational emissions and the high column density
favoring emission from polyatomic molecules that possess rich 
rotational spectra with many lines.

\subsection{Line emission}

In Figure 6, we plot the line luminosities for $^{12}$CO, $^{13}$CO,
and C$^{18}$O, as a
function of upper rotational number, $J_u$.
The results shown in Figure 6 apply to the standard model,
and are presented in tabular form in Table 1. 
Line strengths for rotational transitions of ortho-
and para-H$_2^{16}$O and -H$_2^{18}$O appear in Tables 2 and 3.
Although the abundances of the isotopomers 
$^{13}$CO and C$^{18}$O are assumed to
lie factors of 100 and 500 below the abundance of $^{12}$CO,
the predicted $^{13}$CO and C$^{18}$O
line luminosities are smaller than the corresponding $^{12}$CO
line luminosities by factors of only 2--6 and 5--20 respectively,
a behavior that clearly reflects optical depth effects.
Similarly, the ratio of H$_2^{16}$O to H$_2^{18}$O line
strengths are typically much smaller than the assumed abundance
ratio of 500.
At short wavelengths, corresponding to $J_u$ larger than $\sim 20$
for CO,
the dust radiation field becomes increasingly important and
the rotational transitions are predicted to be {\it absorption} lines. 

Tables 1--3 also show the comparison between the 
line strength predictions of this work and those obtained by NLM
for a highly idealized ``hot core" model.  For many lines the
predicted luminosities are substantially different.
The NLM hot core model differed from the standard model discussed
here in several important respects.  Most importantly, NLM assumed
a constant gas temperature of 100~K, rather than the spatially
varying temperature profile obtained from the self-consistent
calculations reported here. Furthermore, NLM made use 
of an approximate escape probability method rather than an exact 
ALI method in obtaining line strength predictions, and did not include
the interaction between line radiation and infrared dust continuum radiation.
Finally, the assumed outer radius for the NLM was a factor 4 smaller than
that assumed here.

For transitions (such as CO $J=1-0$) that are excited in the very
outer regions of the cloud core, the predictions obtained in this 
study are larger than those of NLM because of the larger outer
radius that we have assumed.  Transitions of
intermediate excitation, however, are typically predicted to show weaker
emission than that predicted by NLM, the temperature in 97\%
of the core's mass being lower than the 100~K value adopted by NLM.
On the other hand, transitions with very high critical densities
(such as the water transitions originating in 
$2_{21}$, $3_{03}$, $3_{12}$ and $3_{21}$ levels) probe the
innermost parts of the cloud core where our model predicts
temperatures {\it larger} than 100~K; for these lines the NLM
model therefore {\it under}predicted the emission line luminosities relative
to the standard model presented here.  

Table 4 shows the predicted luminosities for the far-infrared
fine structure lines of atomic oxygen.  As discussed in \S 5.1 below,
the $\rm ^3P_0 - ^3P_1$ transition near 146 $\mu$m is predicted to
be a strong emission line and the $\rm ^3P_1 - ^3P_2$ transition
near 63 $\mu$m to be a strong absorption line, contrary to
the predictions of the recent model of CHT.

To compare simultaneously the line strengths of the dominant
species, in Figure 7 we plot the spectrum of our standard model for
$^{12}$CO, H$_{2}^{16}$O and OI lines, as well as the continuum.
Here we assume a spectral resolution
$\lambda / \Delta \lambda = 10^{4}$ and a distance to the source of 1 kpc.
Notice that many of the CO lines have large line to continuum flux
ratios.  It is also clear that the low energy water transitions are 
predicted to be in emission, while the high-lying transitions
are in strong absorption against the continuum.    

Our model predicts a population inversion and low-gain maser 
amplification in one transition:
the $3_{13}-2_{20}$ transition of 
water near 183 GHz, which has the largest line/continuum ratio of
any water line.
Low-gain maser emission has indeed been observed in this transition
toward several hot core regions (Cernicharo et al. 1994).
Our model demonstrates that quiescent, radiatively-heated
hot cores are expected to show low-gain maser emission in the
183 GHz $3_{13}-2_{20}$ water line with a luminosity that
can easily account for (in fact, exceed by a factor of $\sim 4$)
that observed toward Orion-KL.  
Further discussion of the 183 GHz and other masing transitions
will appear in 
a future paper (Doty \& Neufeld 1997).

Our use of an ALI method to model the transfer of line radiation
leads straightforwardly to predictions for the spectral line profiles.
In Figures 8--11, we show emergent line profiles
(assuming a beam much larger than the source) for a total of 32 transitions 
of ortho-H$_2^{16}$O, ortho-H$_2^{18}$O, para-H$_2^{16}$O, para-H$_2^{18}$O,
and atomic oxygen.
These figures show the monochromatic luminosity, $L_\nu$,
as a function of Doppler shift from line center.
While some transitions are pure emission or pure absorption
lines, many transitions show the ``self-reversed" (or ``double-peaked")
profiles that are characteristic of line radiation for which the source
function decreases along the line-of-sight towards the observer.
This behavior arises because material of low excitation 
temperature in the outer parts of the cloud core
absorbs line radiation that was emitted by the hot, dense interior.

\subsection{Effect of Model Assumptions}
We have carried out test calculations in order to investigate
the effects of changing some of the physical assumptions
underlying our model. In particular, we varied
the assumed gas density distribution, considered the effect
of H$_2$O desorption from icy mantles above
100K, and varied the assumed
ratio of ortho- to para-H$_2$ as a collision partner with
water.  All these calculations were
carried out for the assumed core mass and source
luminosity of the standard model.

In these test calcuations, the greatest differences
in the predicted line strengths occurred when we changed the 
assumed density distribution from $r^{-2}$ to $r^{-1.5}$
(while leaving the inner
and outer radii and the quantity M$_{0.1}$ unchanged).
For the three CO isotopes, the typical (i.e. median) change
in the predicted line strengths was $\sim 25$ percent,
with the greatest differences occurring for low-lying
transitions.
The typical differences in the water line strengths were
$\sim 40$ percent for
o-H$_{2}^{16}$O and $\sim 70$ percent for p-H$_{2}^{18}$O.
For the two SWAS target lines of water,
the predicted strength of the 557 GHz ortho-water line
increased by 33 percent with the shallower density profile,
while that of the 547 GHz para-water line decreased by a factor
$\sim 6$.  The high-lying water absorption lines were
relatively unaffected, and changes in the predicted
OI line strengths were smaller than 25 percent 
(with the 63 $\mu$m transition still in absorption).

Next, we considered the effect of including desorption
of H$_{2}$O from icy grain mantles above 100K.  
Were carried out test calculations in which, following CHT,
we assumed that 10 percent of the total oxygen was injected
into the gas phase as H$_2$O 
once the grain temperature exceeded 100 K.
This change resulted in relatively small differences in the 
predicted water line strengths, with typical increases of
7, 50, 16, and 39 percent for transitions of
o-H$_{2}^{16}$O, 
o-H$_{2}^{18}$O, p-H$_{2}^{16}$O, and p-H$_{2}^{18}$O respectively. 
No line strength increased by more than a factor 2.4, and the
predicted strengths of the two SWAS water target lines
increased 21 percent (557 GHz) and 34 percent (547 GHz) when the
grain mantle desorption was included.

Finally, we have considered the effect of changing
the ortho- to para-H$_{2}$ ratio to 3:1.  For
most transitions, the differences were minimal,
yielding median differences in the predicted line strengths 
of $\sim 10$ percent.
For the two SWAS water target lines, however, the effects were
atypically large, with the predicted line strengths
increasing by 88 percent (557 GHz) and 77 percent (547 GHz)
when the 3:1 ratio was adopted.

\section{General Parameter Study}

We have extended the calculations described in \S 3 to include a general
parameter study of cloud cores with masses $M_{0.1}$ in the range
$1 - 1000$ M$_\odot$ that contain central sources of luminosity 
$L_*$ in the range $1 - 10^6$ L$_\odot$.  We considered
a total of 21 combinations of $M_{0.1}$ and $L_*$, as specified
in Table 5.

\subsection{Gas and Dust Temperature}

Because the gas and dust temperature distributions
play a large role in determining the emergent spectrum,
we have shown in Figure 12 the gas (solid line) and
dust (dashed-line) temperature distributions for our entire grid of models.
In this figure, each box corresponds to a given value of $M_{0.1}$, while
each set of lines is labeled by ${\rm log}_{10}(L_{*})$.  As in the standard
model, the gas and dust temperatures are always well-coupled in the interior,
but diverge in the exterior of the cloud.  We note also that the
gas and dust temperatures are not necessarily monotonic, due to the 
existence of both internal and external heat sources.

\subsection{Line luminosities}

In Figures 13--17, we present the predicted luminosities for
a total of 44 radio, submillimeter, and far-infrared lines of
$^{12}$CO, ortho-H$_2^{16}$O, ortho-H$_2^{18}$O,
para-H$_2^{16}$O, para-H$_2^{18}$O, and atomic oxygen.
Each box within these figures shows the line
luminosity for a single transition, for each of the 20 models in our grid. The
different symbols correspond to the different valus of $M_{0.1}$:\ open
triangles, open squares, crosses, and open circles represent $M_{0.1}=1,$ $%
10,$ $10^2,$ and $10^3$ M$_{\odot }$ respectively. For each transition, the
upper panel in each full box corresponds to net emission against
the background, while the lower panel corresponds to net absorption. 
The upper and lower panels of each box share the same zero.
In Figure 18, we show the CO rotational lines strengths as
a function of the upper rotational quantum number, $J_u$,
for the case $L_{*}=100$ L$_{\odot }$.
For subset of the models that we considered (with $M_{0.1}=100$ 
M$_{\odot }$ and $L_{*}=100$ L$_{\odot }$)
the data are also presented in tabular form in 
Tables 6 \& 7, along with predictions for rotational lines of
$^{13}$CO and C$^{18}$O.  The entire dataset is available from
the authors in computer-readable form.

The results of our parameter study show several general features.   
Low-lying transitions (viz.\ CO $J=1-0$, CO $J=2-1$,
CO $J=3-2$ and ortho-H$_2$O $1_{10}-1_{01}$) show line luminosities
that are almost independent of the central luminosity, $L_*$,
because they are excited primarily in the outer parts of the cloud
where the heating is dominated by the ISRF rather than by the
central protostar. 
The line luminosities for such transitions
increase roughly as the total assumed area of the cloud core, 
which is proportional to $M_{0.1}^{3/2}$ (c.f.\ Neufeld \& Green 1994).
The line luminosities predicted by our model for low-lying
transitions may be slight underestimates, due to our neglect of
grain photoelectric heating.

In contrast, high-lying transitions show predicted
line strengths that are increasing functions of L$_*$, because
they originate in the cloud interior where the
gas and dust temperatures increase with increasing central 
luminosity.   We expect many such transitions 
to be observed in {\it absorption} against the continuum
radiation emitted by dust, particularly in high-mass cloud cores.
For $^{12}$CO, there is always a value of $J_u$ beyond which the
rotational transitions are all absorption lines: these values are
tabulated in Table 8.  We also note that for certain source
parameters, the CO line luminosities can be a rather complex
function of $J_u$, with line emission predicted for 
low-$J_u$ transitions, line absorption for intermediate-$J_u$
transitions, line emission again for transitions of 
yet higher $J_u$, and finally absorption again for the
highest $J_u$ transitions.  This behavior, which is apparent
in Figure 18, can occur when the gas
temperature profile is non-monotonic.

\section{Discussion}

\subsection{Comparison with previous studies}

The main differences between the model presented here and the
highly-idealized hot core models of NLM have been discussed
in \S 3.5 above.  Despite the quantitative differences between
the results obtained from the two models, the present study confirms the
{\it qualitative} result of NLM that far-infrared and submillimeter
lines dominate the emission from gas in dense molecular cloud
cores.  

Another detailed model for the thermal and chemical
structure within -- and line emission  from -- dense molecular cloud 
cores was presented 
recently by CHT, for the case of cores containing low-mass protostars
of luminosity 20 or 65 L$_\odot$.  Our present study
greatly expands the range of $L_*$ and $M_{0.1}$
considered, and makes use of a ALI radiative transfer method 
rather than the approximate escape probability approach adopted
by CHT.  On the other hand, the model of CHT included
four effects that are not considered in the present work.
First, CHT treated the cloud chemistry in a time-dependent
manner, following the chemical evolution of water
and other species after they vaporized from icy grain mantles 
(c.f.\ \S2.2 above.).  Second, CHT considered the effects
of velocity gradients in the cloud core, using the
collapse solution of Shu (1977) to specify the evolution
of the density profile and infall velocities.
Third, CHT found that compressional heating can be important in
the core interior; we do not include this process as we do
not include the infall of material.
Finally, CHT included vibrational pumping by near-infrared photons,
a process that we have neglected. In the static case 
that we have considered, the vibrational pumping rate will
be significantly dimished by radiative trapping, in contrast
to the infall case considered by CHT where the pumping
photons could propagate due to a spatial velocity gradient.

In general, our study agrees with many of the 
qualitative results of CHT, {\it except in the prediction of 
{\rm OI}
$\rm ^3P_1 - ^3P_2$ line emission}.  In particular, while CHT predicted
that this line should be a strong emission line, we find that
the $\rm ^3P_1 - ^3P_2$ OI line
should be in absorption for a wide range of source parameters.
This difference arises because the temperature profile in the CHT
model is higher due to their inclusion of compressional heating.  In 
our model, the predicted absorption is also enhanced due to our more
extended envelopes and our assumption of constant line widths.

As a check, we have used our ALI code to determine the expected 
line strengths for a cloud core with the same temperature profile,
density profile, and linewidths assumed by CHT.  
We find the OI 63 $\mu$m line to be in {\it emission} for that case,
deriving fluxes for this and other lines that agree to within a 
factor of 1.5 - 5 with the CHT results.  The CHT predictions
are consistently larger than our corresponding ALI results, a
behaviour which -- we believe -- results from CHT's use of an
approximate escape probability method for the line transfer.

Although beyond the scope of this paper, it is interesting to 
consider the possible implications of the transition of the 
$\rm ^3P_1 - ^3P_2$ OI line from emission (CHT) to absorption
(this study).  In particular, over a wide  range of parameters
it appears that quiescent sources without the infall of material
are predicted to 
exhibit absorption in this line, while collapsing sources are
predicted to show
net emission.  Further study on the effects of a velocity gradient
and compressional heating are needed to understand the range
of applicability of this result.  If the result is robust, then
the simple presence of emission in the 63 $\mu$m OI
line might serve as a collapse signature.

\subsection{Observational implications: far-IR and submillimeter lines}

Many of the
far-infrared and submillimeter transitions
that we expect to be characteristic of dense molecular cloud cores
lie at wavelengths for which ground-based observations
are made impossible by the effects of atmospheric absorption.
NLM have emphasized the crucial role that airborne and satellite 
observatories will play in testing our models for dense cloud cores; 
these observatories include the recently-launched
{\it Infrared Space Observatory} (ISO),
the {\it Submillimeter Wave Astronomy Satellite} (SWAS, scheduled
for launch in 1997), and the planned {\it Stratospheric Observatory 
for Infrared Astronomy} (SOFIA).

ISO, SWAS and SOFIA have complementary observational capabilities.
ISO provides an ideal
platform for the study of far-infrared transitions of H$_2^{16}$O
and high-$J$ ($J_u > 13$) CO transitions, although at a spectral resolution
($\sim 30\rm \, km\,s^{-1}$ in the 45 -- 180 $\mu$m range)
that will not allow the line profiles from quiescent cores to
be resolved.  SWAS, by contrast,
will carry out high spectral resolution ($\sim 0.6 \rm \, km\,s^{-1}$)
observations of
5 spectral lines in the 487 -- 557 GHz
frequency range: the $1_{10}- 1_{01}$ lines of H$_2^{16}$O and
H$_2^{18}$O; the $\rm ^3P_1 - ^3P_0$ fine structure line of atomic carbon;
the $J=5-4$ line of $^{13}$CO, and the $3(3)-1(2)$ line of molecular oxygen.
With a primary mirror
of diameter 4 times as large as that of either ISO or SWAS, SOFIA
will provide much better {\it spatial} resolution than any previous airborne
or infrared satellite observatory, and with appropriate instrumentation will
cover much of the infrared and submillimeter spectral region.  However,
even at a flying altitude of 14 km, atmospheric attenuation will
severely limit SOFIA's ability to observe water or molecular oxygen.
For all three observatories, detailed cloud core models such as those
presented in this study will be essential in deriving molecular 
abundances from the line strengths that are observed.

Finally, we note that many transitions that
are unobservable from the ground for Galactic sources 
are shifted into atmospheric
windows for high redshift galaxies.  Thus far, CO rotational
emissions have been detected unequivocally from three high redshift
systems: the Cloverleaf (at z = 2.551;  
Barvainis, et al. 1992, Barvainis, et al. 1994,
Hewitt \& Burbidge 1989),
the ultraluminous IR galaxy F10214+4724 (at z = 2.286;
Brown \& Vanden Bout 1991,   
Downes, Solomon, \& Radford 1993) and a companion (a gaseous nebula at
z=4.702; Petitjean, et al. 1996) to the
quasar BR1202-0725.  Searches for water in these
sources would provide a valuable probe of the chemistry and
molecular excitation at high redshift.  Prime target lines for these sources
are listed in Table 9, along with their appropriate redshift.
In the two nearest
sources, the $2_{11}-2_{02}$ water line,  
redshifted to 228.9 GHz in F10214+4724 and to 211.8 GHz in the Cloverleaf,
should be the strongest observable line.
In the companion to BR1202-0725,
the strongest line should be the $3_{21}-3_{12}$
water line (redshifted to 203.9 GHz), with 
many others
also expected to be strong -- a prediction which is robust
for a relatively wide range of model parameters.

\subsection{Observational implications: mid-IR vibrational absorption bands}

Although the present study has emphasized molecular rotational
lines and atomic fine structure lines at submillimeter
and far-infrared wavelengths, ISO also affords us the opportunity
of studying vibrational absorption bands in the near- and mid-infrared
spectral region.  In particular, recent ISO observations by 
Helmich et al. (1996), and by van Dishoeck 
\& Helmich (1996, hereafter vDH) have led to the detection
of the $\nu_2$ vibrational band of water vapor near 6 $\mu$m
in absorption toward several protostars.
 
For the source GL 2591, vDH estimated the water abundance
by assuming that the absorption takes place in an
isothermal region that lies entirely in front of --
and completely covers -- the source of 6 $\mu$m
continuum emission.  That assumption leads to an
estimate of $2 \times 10^{18}\rm \, cm^{-2}$ for the
water column density.  Given a CO column density that
has been estimated by a similar method as $2 \times 10^{19} \rm \, cm^{-2}$
(Mitchell et al.\ 1989, 1990), and an assumed CO abundance relative
to H$_2$ of $2 \times 10^{-4}$, vDH derived a mean water vapor abundance
of $2 \times 10^{-5}$ along the line-of-sight to GL 2591.

In our cloud core models, the predicted strength of the $\nu_2$
absorption band is complicated by several effects.
First, the model predicts strong spatial variations in the water abundance.
Second, the water vapor that is responsible for the $\nu_2$ band absorption
is largely cospatial with the dust that gives rise to the
6 $\mu$m continuum emission.  Third, the source is assumed to have
a spherical (rather than a plane-parallel) geometry.

For comparison with the water vapor column density inferred
from the simpler analysis of vDH, we have computed an {\it effective}
water column density for our cloud core models.  To do so, we defined
an effective photosphere for the 6 $\mu$m radiation, and
computed the beam-averaged (and intensity-weighted) water
column density from the cloud surface to that photosphere.
Table 10 shows the effective column densities thereby derived
for the $\nu_2$ band of water, as well as for the $v=2-0$ band of CO.
Given estimates of $\sim 50 \rm \,M_{\odot}$ for $M_{0.1}$ in GL 2591
(Carr et al.\ 1995)
and $2 \times 10^{4}$ L$_{\odot}$ for $L_*$
(Mozurkewich, Schwartz, \& Smith 1986), the results given in 
Table 10 are in acceptable agreement with the column densities
derived from observations of GL 2591.  Since our model does not
include the vaporization of water from icy grain mantles, we
conclude that high-temperature gas-phase chemistry in the
interior of the cloud core is adequate by itself 
to explain the relatively large average water abundances derived 
by vDH.  The importance of high-temperature chemistry is further
supported by the relatively high excitation temperature ($\sim 300$ K)
inferred by vDH, although more detailed modeling will clearly be required
to obtain quantitative predictions against which the excitation
temperature can be compared.

\section{Summary}

We have constructed a detailed, self-consistent model for
the thermal balance, chemistry, and continuum and line
radiative transfer in dense molecular cloud cores
that contain a central protostar.  Our model
makes specific predictions about the temperature and 
molecular abundances as a function of radial position
within such a cloud core.  We have carried out a general
parameter study to obtain predictions
for the spectra of molecular cloud cores as a function
of the cloud core mass and the luminosity of
the central luminosity source.  The principal results of our
study may be summarized as follows:
{\parindent 0pt

1. The gas and dust temperatures are well-coupled in the dense
interiors of molecular cloud cores.  In the outer regions of
cloud cores, however, the
gas cools efficiently through CO rotational lines and
the gas temperature drops significantly below the dust temperature.  
Furthermore, the temperature distributions need not be monotonic 
due to the existence of both internal and external heat sources.

2. Due to the existence of both temperature and density gradients,
the molecular abundances are not expected to be constant
within dense molecular cloud cores.  In
particular, the predicted H$_{2}$O abundance varies by over a factor of
1000 in our models.  

3. We have obtained predictions for the strengths of
a large number of transitions of $^{12}$CO, $^{13}$CO, C$^{18}$O,
ortho- and para-H$_{2}^{16}$O, ortho- and para-H$_{2}^{18}$O,
and atomic oxygen.  In general, we find that the lowest lying transitions
are generally emission lines with strengths that are almost 
independent of the central luminosity.
Higher lying transitions, by contrast, generally show
line strengths that are proportional
to the central luminosity and are 
often predicted to be absorption lines.

4. Many of the most prominent lines expected in the spectra 
of dense molecular cloud cores are unobservable from 
ground-based telescopes but should be readily detectable
from airborne and space-based observatories such as
ISO, SWAS and SOFIA.
The $\rm ^3P_1 - ^3P_2$ fine structure line of atomic oxygen 
near 63 $\mu$m is expected to be in strong absorption for many 
quiescent sources.
Quiescent, radiatively-heated hot cores are expected to 
show low-gain maser emission in the 183 GHz $3_{13}-2_{20}$
water line, such as has been observed toward several
hot core regions using ground-based telescopes.

5. Many lines are predicted to show the ``self-reversed" or
``double-peaked" profiles that are characteristic
of line emission from a source in which the excitation
temperature decreases along the line-of-sight toward the observer.

6. Our model successfully accounts for the water column
densities inferred by van Dishoeck and Helmich from
observations of the $\nu_2$ absoprtion band of water toward
the source GL 2591.}
 
{\acknowledgments
We thank Ali Yazdanfar for his help in calculating the thermal 
balance in the gas.  We are grateful to 
David Hollenbach and to
Cecilia Ceccarelli for helpful comments.
We acknowledge with gratitude the support of NASA grant NAGW-3147
from the Long Term Space Astrophysics Research Program and of
a NSF Young Investigator Award to D.A.N.
}

\makeatletter

\ifx\revtex@jnl\jnl@aj\let\tablebreak=\nl\fi
\makeatother

\newpage

\begin{table}
\tablewidth{40pc}
\begin{center}
\caption{CO line luminosities for various transitions
in the standard model.}
\vspace{0.3cm}
\begin{tabular}{rrrrrrr} \hline \hline
{} &
$^{12}$CO &
$^{12}$CO &
$^{13}$CO &
$^{13}$CO &
C$^{18}$O &
C$^{18}$O \\
{} &
This Work & 
NLM &
This Work & 
NLM & 
This Work &
NLM \\

\hline

{J=1$\rightarrow $0} & $4.3(-3)$ & $1.8(-3)$ & $2.3(-3)$ & 
$2.5(-4)$ & $7.1(-4)$ & $6.2(-5)$ \nl
{2$\rightarrow $1} & $2.1(-2)$ & $1.5(-2)$ & $1.4(-2)$ & $%
5.0(-3)$ & $5.0(-3)$ & $1.6(-3)$ \nl
{3$\rightarrow $2} & $4.9(-2)$ & $5.1(-2)$ & $2.9(-2)$ & $%
2.3(-2)$ & $1.1(-2)$ & $8.6(-3)$ \nl
{4$\rightarrow $3} & $8.8(-2)$ & $1.2(-1)$ & $4.1(-2)$ & $%
5.7(-2)$ & $1.5(-2)$ & $2.3(-2)$ \nl
{5$\rightarrow $4} & $1.3(-1)$ & $2.2(-1)$ & $4.9(-2)$ & $%
1.1(-1)$ & $1.8(-2)$ & $4.2(-2)$ \nl
{6$\rightarrow $5} & $1.6(-1)$ & $3.8(-1)$ & $5.3(-2)$ & $%
1.5(-1)$ & $2.0(-2)$ & $5.8(-2)$ \nl
{7$\rightarrow $6} & $1.9(-1)$ & $5.7(-1)$ & $5.2(-2)$ & $%
1.9(-1)$ & $1.9(-2)$ & $6.3(-2)$ \nl
{8$\rightarrow $7} & $1.9(-1)$ & $8.2(-1)$ & $4.9(-2)$ & $%
2.0(-1)$ & $1.8(-2)$ & $6.2(-2)$ \nl
{9$\rightarrow $8} & $1.9(-1)$ & $1.1(-0)$ & $4.3(-2)$ & $%
1.8(-1)$ & $1.5(-2)$ & $5.3(-2)$ \nl
{10$\rightarrow $9} & $1.7(-1)$ & $1.4(-0)$ & $3.6(-2)$ & $%
1.5(-1)$ & $1.3(-2)$ & $4.2(-2)$ \nl
{11$\rightarrow $10} & $1.5(-1)$ & $1.5(-0)$ & $3.0(-2)$ &
$1.1(-1)$ & $1.0(-2)$ & $3.1(-2)$ \nl
{12$\rightarrow $11} & $1.1(-1)$ & $1.3(-0)$ & $2.2(-2)$ & 
$6.7(-2)$ & $7.5(-3)$ & $1.8(-2)$ \nl
{13$\rightarrow $12} & $8.8(-2)$ & $1.1(-0)$ & $1.7(-2)$ & 
$4.4(-2)$ & $5.7(-3)$ & $1.1(-2)$ \nl
{14$\rightarrow $13} & $6.8(-2)$ & $7.9(-1)$ & $1.3(-2)$ & 
$2.7(-2)$ & $4.1(-3)$ & $6.4(-3)$ \nl
{15$\rightarrow $14} & $5.1(-2)$ & $5.2(-1)$ & $9.3(-3)$ &
$1.5(-2)$ & $2.9(-3)$ & $3.5(-3)$ \nl
{16$\rightarrow $15} & $3.7(-2)$ & $3.2(-1)$ & $6.6(-3)$ &
$8.1(-3)$ & $1.9(-3)$ & $1.7(-3)$ \nl
{17$\rightarrow $16} & $2.4(-2)$ & $1.9(-1)$ & $4.4(-3)$ & 
$3.9(-3)$ & $1.2(-3)$ & $8.1(-4)$ \nl
\\
\hline
\end{tabular}
\end{center}
Note. --- In this table, $a(b)$ means $a \times 10^{b}$ L$_{\odot}$
\label{tbl1}
\end{table}

\clearpage

\makeatletter

\ifx\revtex@jnl\jnl@aj\let\tablebreak=\nl\fi
\makeatother

\begin{table}
\tablewidth{40pc}
\begin{center}
\caption{
Ortho-H$_{2}$O line luminosities for various transitions
in the standard model.}
\vspace{0.3cm}
\begin{tabular}{rrrrr} \hline \hline
{} &
{o-H$_{2}^{16}$O} &
{o-H$_{2}^{16}$O} &
{o-H$_{2}^{18}$O} &
{o-H$_{2}^{18}$O} \\
{} &
{This Work} & 
{NLM} &
{This Work} & 
{NLM}
\\ \hline

$1_{10} \rightarrow 1_{01}$ & $3.3(-3)$ & $1.8(-2)$ & $4.1(-4)$ &
    $4.6(-4)$ \nl
$3_{12} \rightarrow 3_{03}$ & $1.4(-1)$ & $1.1(-2)$ & $1.1(-2)$ &
    N/A \nl
$3_{12} \rightarrow 2_{21}$ & $1.3(-1)$ & $2.1(-2)$ & $3.8(-3)$ &
    N/A \nl
$3_{21} \rightarrow 3_{12}$ & $3.0(-1)$ & $8.7(-3)$ & $1.5(-2)$ &
    N/A \nl
$2_{21} \rightarrow 2_{12}$ & $2.1(-2)$ & $4.1(-2)$ & $-9.7(-3)$ &
    $6.0(-4)$ \nl
$2_{12} \rightarrow 1_{01}$ & $-3.5(-1)$ & $2.0(-1)$ & $-6.4(-2)$ &
     $3.9(-3)$ \nl
$3_{03} \rightarrow 2_{12}$ & $-2.5(-2)$ & $9.2(-2)$ & $-3.2(-3)$ &
    $1.7(-3)$ \nl
$2_{21} \rightarrow 1_{10}$ & $-1.2(-0)$ & N/A & $-9.1(-2)$ & N/A \nl
$3_{21} \rightarrow 2_{12}$ & $-2.4(-0)$ & N/A & $-1.3(-1)$ & N/A \nl
\\
\hline
\end{tabular}
\end{center}
Note. --- In this table, $a(b)$ means $a \times 10^{b}\rm \,L_\odot$.
\label{tbl2}
\end{table}

\clearpage

\begin{table}
\tablewidth{40pc}
\begin{center}
\caption{
Para-H$_{2}$O line luminosities for various transitions
in the standard model.}
\vspace{0.3cm}
\begin{tabular}{rrrrr} \hline \hline
{} &
p-H$_{2}^{16}$O &
p-H$_{2}^{16}$O &
p-H$_{2}^{18}$O &
p-H$_{2}^{18}$O \\
{} &
This Work & 
NLM &
This Work & 
NLM \\
\hline

$3_{13} \rightarrow 2_{20}$ & $6.8(-4)$ & N/A & $3.0(-5)$ & N/A \nl
$2_{11} \rightarrow 2_{02}$ & $1.1(-1)$ & $1.3(-2)$ & $5.0(-3)$ &
    N/A \nl
$2_{02} \rightarrow 1_{11}$ & $1.9(-1)$ & $4.9(-2)$ & $1.1(-2)$ &
    $8.4(-4)$ \nl
$1_{11} \rightarrow 0_{00}$ & $-7.1(-2)$ & $7.6(-2)$ & $-7.5(-3)$ &
    $1.5(-3)$ \nl
$2_{20} \rightarrow 2_{11}$ & $2.6(-1)$ & $8.6(-3)$ & $9.7(-3)$ &
    N/A \nl
$3_{22} \rightarrow 3_{13}$ & $3.6(-2)$ & $1.3(-2)$ & $-4.8(-3)$ &
     N/A \nl
$3_{13} \rightarrow 2_{02}$ & $-2.6(-1)$ & N/A & $-2.0(-2)$ & N/A \nl
$2_{20} \rightarrow 1_{11}$ & $-1.6(-0)$ & N/A & $-8.1(-2)$ & N/A \nl
$3_{22} \rightarrow 2_{11}$ & $-1.2(-0)$ & N/A & $-3.8(-2)$ & N/A \nl
\\
\hline
\end{tabular}
\end{center}
Note. --- In this table, $a(b)$ means $a \times 10^{b}\rm \,L_\odot$.
\label{tbl3}
\end{table}

\clearpage

\begin{table}
\begin{center}
\caption{
OI line luminosities for various transitions
in the standard model.}
\vspace{0.3cm}
\begin{tabular}{rrr} \hline \hline
{} &
$^{16}$O &
$^{16}$O \\ 
{} &
This Work & 
NLM \\
\hline

J=$0 \rightarrow 1$ & $9.0(-2)$ & N/A \nl
J=$1 \rightarrow 2$ & $-1.4(-0)$ & N/A  \nl

\\
\hline
\end{tabular}
\end{center}
Note. --- In this table, $a(b)$ means $a \times 10^{b}\rm \,L_\odot$.
\label{tbl3b}
\end{table}

\clearpage

\begin{table}
\tablewidth{40pc}
\begin{center}
\caption{
Grid of models considered in this work.  ``X" denotes
models run.
}
\vspace{0.3cm}
\begin{tabular}{lccccccc} \hline \hline

$M_{0.1}/$M$_{\odot}$ &
$1$ L$_{\odot}$ &
$10$ L$_{\odot}$ &
$10^{2}$ L$_{\odot}$ &
$10^{3}$ L$_{\odot}$ &
$10^{4}$ L$_{\odot}$ &
$10^{5}$ L$_{\odot}$ &
$10^{6}$ L$_{\odot}$
\\
\hline

$1$ & X & X & X & --- & --- & --- & --- \\ 
$10$ & X & X & X & X & --- & --- & --- \\ 
$60$ & --- & --- & --- & --- & --- & Std. Model & --- \\ 
$10^2$ & X & X & X & X & X & X & --- \\ 
$10^3$ & X & X & X & X & X & X & X \\ 
\hline \hline
\end{tabular}
\end{center}
\label{tbl4}
\end{table}

\clearpage

\begin{deluxetable}{rrrrrr}
\tablewidth{39pc}
\tablecaption{Background-subtracted line luminosities}
\tablehead{
\colhead{Transition} &
\colhead{Frequency (GHz)} &
\colhead{$1$ M$_{\odot}$} &
\colhead{$10$ M$_{\odot}$} &
\colhead{$10^{2}$ M$_{\odot}$} &
\colhead{$10^{3}$ M$_{\odot}$}
}
\startdata
\multicolumn{1}{l}{$^{12}$CO} &  &  &  &  &   \nl
J=1$\rightarrow $0 &  $115.271$ & $2.3(-6)$ & $1.1(-4)$ & $5.2(-3)$ & $2.1(-1)$ \nl
$4 \rightarrow $3 &   $461.041$ & $5.1(-5)$ & $9.0(-4)$ & $1.5(-2)$ & $2.0(-1)$  \nl
$7 \rightarrow $6 &   $806.652$ & $1.1(-4)$ & $5.0(-4)$ & $1.2(-3)$ & $-7.9(-4)$ \nl
$10 \rightarrow $9 &  $1151.985$ & $8.6(-5)$ & $2.4(-4)$ & $-3.5(-5)$ & $-1.2(-4)$ \nl
$13 \rightarrow $12 & $1496.923$ & $4.6(-5)$ & $6.2(-5)$ & $-7.0(-5)$ & $-1.1(-6)$ \nl
$16 \rightarrow $15 & $1841.346$ & $2.4(-5)$ & $8.4(-6)$ & $-1.0(-5)$ & $-3.7(-8)$ \nl
$19 \rightarrow $18 & $2185.135$ & $1.3(-5)$ & $1.4(-7)$ & $-2.1(-7)$ & $-3.4(-8)$ \nl
$22 \rightarrow $21 & $2528.172$ & $6.9(-6)$ & $1.7(-6)$ & $3.7(-9)$ & $-2.0(-8)$ \nl
 & & & & & \nl
\multicolumn{1}{l}{C$^{18}$O} & &  &  & &  \nl
J=1$\rightarrow $0 &  $109.782$ & $9.6(-7)$ & $3.2(-5)$ & $1.1(-3)$ & $3.1(-2)$ \nl
$4 \rightarrow $3 &   $439.089$ & $1.4(-5)$ & $1.7(-4)$ & $1.1(-3)$ & $2.4(-3)$ \nl
$7 \rightarrow $6 &   $768.252$ & $1.3(-5)$ & $7.2(-5)$ & $1.2(-4)$ & $-7.2(-5)$ \nl
$10 \rightarrow $9 &  $1097.164$ & $7.5(-6)$ & $2.4(-5)$ & $3.1(-6)$ & $-1.4(-6)$ \nl
$13 \rightarrow $12 & $1425.718$ & $3.7(-6)$ & $7.3(-6)$ & $-2.9(-8)$ & $-5.0(-8)$ \nl
$16 \rightarrow $15 & $1753.805$ & $1.6(-6)$ & $1.7(-6)$ & $7.1(-9)$ & $-3.6(-8)$ \nl
$19 \rightarrow $18 & $2081.311$ & $5.0(-7)$ & $2.7(-7)$ & $2.6(-10)$ & $-3.4(-8)$ \nl
$22 \rightarrow $21 & $2408.140$ & $1.3(-7)$ & $3.0(-8)$ & $-7.8(-11)$ & $-2.0(-8)$ \nl
 & & & & & \nl
\multicolumn{1}{l}{$^{13}$CO} &  & &  & &  \nl
J=1$\rightarrow $0 &  $110.201$ & $2.1(-6)$ & $7.2(-5)$ & $2.6(-3)$ & $7.9(-2)$ \nl
$4 \rightarrow $3 &   $440.765$ & $3.1(-5)$ & $3.8(-4)$ & $2.4(-3)$ & $6.8(-3)$ \nl
$7 \rightarrow $6 &   $771.185$ & $3.2(-5)$ & $1.6(-4)$ & $2.7(-5)$ & $-2.1(-4)$ \nl
$10 \rightarrow $9 &  $1101.353$ & $1.8(-5)$ & $5.1(-5)$ & $-4.9(-6)$ & $-6.2(-6)$ \nl
$13 \rightarrow $12 & $1431.160$ & $9.2(-6)$ & $1.5(-5)$ & $-2.4(-6)$ & $-6.2(-8)$ \nl
$16 \rightarrow $15 & $1760.499$ & $4.8(-6)$ & $5.0(-6)$ & $5.0(-9)$ & $-3.6(-8)$ \nl
$19 \rightarrow $18 & $2089.240$ & $2.0(-6)$ & $1.2(-6)$ & $1.7(-9)$ & $-3.4(-8)$ \nl
$22 \rightarrow $21 & $2417.308$ & $5.9(-7)$ & $1.4(-7)$ & $-4.3(-11)$ & $-2.0(-8)$ \nl
 &  & & &  & \nl
\multicolumn{1}{l}{o-H$_{2}^{16}$O} &  & &  & &  \nl
$1_{10} \rightarrow 1_{01}$ & $556.936$ & $9.2(-6)$ & $1.5(-4)$ & $1.1(-3)$ &
    $-8.6(-3)$ \nl
$3_{12} \rightarrow 3_{03}$ & $1097.365$ & $1.9(-5)$ & $4.6(-5)$ & $-5.0(-4)$ &
    $-6.6(-4)$ \nl
$3_{12} \rightarrow 2_{21}$ & $1153.127$ & $2.5(-5)$ & $8.1(-5)$ & $-2.4(-4)$ &
    $-2.4(-4)$ \nl
$3_{21} \rightarrow 3_{12}$ & $1162.911$ & $3.4(-5)$ & $1.2(-4)$ & $-1.3(-4)$ &
    $-1.4(-4)$ \nl
$2_{21} \rightarrow 2_{12}$ & $1661.008$ & $4.3(-5)$ & $-1.1(-4)$ & $-2.1(-3)$ &
    $5.4(-3)$ \nl
$2_{12} \rightarrow 1_{01}$ & $1669.905$ & $8.0(-7)$ & $-5.7(-4)$ & $-5.3(-3)$ &
    $-1.1(-2)$ \nl
$3_{03} \rightarrow 2_{12}$ & $1716.770$ & $2.8(-5)$ & $-2.7(-4)$ & $-2.8(-3)$ &
    $-1.4(-3)$ \nl
$2_{21} \rightarrow 1_{10}$ & $2773.978$ & $-1.2(-4)$ & $-1.4(-3)$ & $-2.5(-3)$ &
    $-9.6(-3)$ \nl
$3_{21} \rightarrow 2_{12}$ & $3977.047$ & $-2.7(-4)$ & $-1.1(-3)$ & $-2.6(-4)$ &
   $-3.4(-5)$ \nl
 & &  & &  & \nl
\multicolumn{1}{l}{o-H$_{2}^{18}$O} & & &  & &  \nl
$1_{10} \rightarrow 1_{01}$ & $547.676$ & $6.6(-7)$ & $1.1(-5)$ & $-2.5(-5)$ &
    $-5.1(-3)$ \nl
$3_{12} \rightarrow 3_{03}$ & $1095.627$ & $3.4(-6)$ & $1.9(-5)$ & $-4.3(-6)$ &
    $-8.5(-6)$ \nl
$3_{12} \rightarrow 2_{21}$ & $1181.394$ & $1.6(-6)$ & $5.0(-6)$ & $-8.7(-7)$ &
    $-6.5(-7)$ \nl
$3_{21} \rightarrow 3_{12}$ & $1136.704$ & $3.9(-6)$ & $2.4(-5)$ & $1.2(-5)$ &
    $-1.1(-6)$ \nl
$2_{21} \rightarrow 2_{12}$ & $1633.484$ & $4.2(-6)$ & $1.1(-5)$ & $-1.3(-4)$ &
    $-2.2(-5)$ \nl
$2_{12} \rightarrow 1_{01}$ & $1655.868$ & $-5.5(-7)$ & $-1.9(-4)$ & $-2.6(-3)$ &
    $-3.4(-3)$ \nl
$3_{03} \rightarrow 2_{12}$ & $1719.250$ & $5.0(-6)$ & $-2.6(-5)$ & $-3.2(-4)$ &
    $-4.8(-5)$ \nl
$2_{21} \rightarrow 1_{10}$ & $2741.675$ & $-1.4(-5)$ & $-3.3(-4)$ & $-7.3(-4)$ &
    $-2.7(-5)$ \nl
$3_{21} \rightarrow 2_{12}$ & $3951.581$ & $-2.5(-5)$ & $-1.5(-4)$ & $-2.1(-5)$ &
   $-2.7(-8)$  \nl
 & & & &  & \nl
\multicolumn{1}{l}{p-H$_{2}^{16}$O} & & &  & &  \nl
$3_{13} \rightarrow 2_{20}$ & $183.310$ & $2.3(-7)$ & $1.2(-6)$ & $5.5(-6)$ &
    $2.4(-5)$ \nl
$2_{11} \rightarrow 2_{02}$ & $752.033$ & $2.5(-5)$ & $1.5(-4)$ & $1.1(-4)$ &
    $-2.0(-3)$ \nl
$2_{02} \rightarrow 1_{11}$ & $987.927$ & $4.4(-5)$ & $2.0(-4)$ & $-8.2(-4)$ &
    $-6.5(-3)$ \nl
$1_{11} \rightarrow 0_{00}$ & $1113.344$ & $1.4(-5)$ & $-6.4(-5)$ & $-3.8(-3)$ &
    $-4.0(-2)$ \nl
$2_{20} \rightarrow 2_{11}$ & $1228.789$ & $3.5(-5)$ & $1.2(-4)$ & $-5.5(-4)$ &
    $-4.7(-4)$ \nl
$3_{22} \rightarrow 3_{13}$ & $1919.360$ & $2.0(-5)$ & $-7.7(-5)$ & $-5.1(-4)$ &
    $-1.2(-5)$ \nl
$3_{13} \rightarrow 2_{02}$ & $2164.132$ & $2.6(-6)$ & $-5.5(-4)$ & $-2.3(-3)$ &
    $-2.2(-4)$ \nl
$2_{20} \rightarrow 1_{11}$ & $2968.750$ & $-1.5(-4)$ & $-1.3(-3)$ & $-1.6(-3)$ &
    $-5.9(-4)$ \nl
$3_{22} \rightarrow 2_{11}$ & $3331.459$ & $-1.4(-4)$ & $-9.7(-4)$ & $-4.8(-4)$ &
    $-2.8(-7)$ \nl
 & & & &  & \nl
\multicolumn{1}{l}{p-H$_{2}^{18}$O} & & &  & &  \nl
$3_{13} \rightarrow 2_{20}$ & $203.408$ & $8.3(-10)$ & $2.1(-9)$ & $2.2(-8)$ &
    $1.2(-7)$ \nl
$2_{11} \rightarrow 2_{02}$ & $745.320$ & $1.7(-6)$ & $2.2(-5)$ & $6.2(-5)$ &
    $-7.3(-5)$ \nl
$2_{02} \rightarrow 1_{11}$ & $994.675$ & $4.0(-6)$ & $3.9(-5)$ & $-4.4(-6)$ &
    $-4.0(-2)$ \nl
$1_{11} \rightarrow 0_{00}$ & $1101.698$ & $1.9(-6)$ & $-2.0(-5)$ & $-1.3(-3)$ &
    $-8.7(-3)$ \nl
$2_{20} \rightarrow 2_{11}$ & $1199.006$ & $3.1(-6)$ & $3.0(-5)$ & $3.6(-5)$ &
    $-4.5(-6)$ \nl
$3_{22} \rightarrow 3_{13}$ & $1894.324$ & $2.9(-6)$ & $4.9(-6)$ & $-4.9(-6)$ &
    $-1.3(-7)$ \nl
$3_{13} \rightarrow 2_{02}$ & $2147.733$ & $2.0(-6)$ & $-4.3(-5)$ & $-1.5(-4)$ &
    $-3.2(-6)$ \nl
$2_{20} \rightarrow 1_{11}$ & $2939.000$ & $-1.1(-5)$ & $-2.1(-4)$ & $-2.8(-4)$ &
    $-1.8(-6)$ \nl
$3_{22} \rightarrow 2_{11}$ & $3296.737$ & $-5.2(-6)$ & $-4.8(-5)$ & $-9.8(-6)$ &
    $-8.9(-9)$  \nl
 & & & &  & \nl
\multicolumn{1}{l}{${}^{16}$O} &  & & & &  \nl
$0 \rightarrow 1$ & $2060.068$ & $4.6(-5)$ & $2.6(-5)$ & $-2.3(-5)$ &
    $-5.9(-8)$ \nl
$1 \rightarrow 2$ & $4745.804$ & $-1.8(-4)$ & $-6.8(-4)$ & $-5.7(-5)$ &
    $2.7(-4)$ \nl
 & & & &  & \nl
\enddata
\tablecomments{Background-subtracted line luminosities for the
subset of models with $L_{*}=100$ L$_\odot$.
In this table, $a(b)$ means $a \times 10^{b}$ L$_\odot$.}
\label{tbl6}
\end{deluxetable}

\clearpage

\begin{deluxetable}{rrrrrrr}
\tablewidth{40pc}
\tablecaption{
Background-subtracted line luminosities}
\tablehead{
{Species} & & & & & & \\ 
{Transition} &
{$1$ L$_{\odot}$} &
{$10$ L$_{\odot}$} &
{$10^{2}$ L$_{\odot}$} &
{$10^{3}$ L$_{\odot}$} &
{$10^{4}$ L$_{\odot}$} &
{$10^{5}$ L$_{\odot}$} 
}

\startdata

\multicolumn{1}{l}{$^{12}$CO} &  &  & & & &  \nl
J=1$\rightarrow $0 & $5.2(-3)$ & $5.2(-3)$ & $5.2(-3)$ & {$5.3(-3$)} &
 {$5.9(-3)$} & $8.9(-3)$ \nl
 $4 \rightarrow $3 & $1.2(-2)$ & $1.3(-2)$ & $1.5(-2)$ & {$2.8(-2)$} &
 {$7.5(-2)$} & $1.7(-1)$ \nl
 $7 \rightarrow $6 & $-1.0(-6)$ & $3.9(-5)$ & $1.2(-3)$ & {$1.4(-2)$} &
 {$8.9(-2)$} & $3.3(-1)$ \nl
 $10 \rightarrow $9 & $8.7(-7)$ & $-1.1(-5)$ & $-3.5(-5)$ & {$2.5(-3)$} &
 {$3.8(-2)$} & $2.5(-1)$ \nl
 $13 \rightarrow $12 & $4.8(-9)$ & $2.5(-7)$ & $-7.0(-5)$ & {$-5.1(-4)$} &
 {$6.1(-3)$} & $9.9(-2)$ \nl
 $16 \rightarrow $15 & $-8.6(-11)$ & $3.0(-8)$ & $-1.0(-5)$ & {$-4.4(-4)$} &
 {$-2.6(-3)$} & $2.4(-2)$ \nl
 $19 \rightarrow $18 & $-9.4(-11)$ & $-3.6(-11)$ & $-2.1(-7)$ & {$-1.2(-4)$} &
 {$-2.6(-3)$} & $-7.8(-3)$ \nl
 $22 \rightarrow $21 & $-6.1(-11)$ & $-6.2(-11)$ & $3.7(-9)$ & {$-1.5(-5)$} &
 {$-1.0(-5)$} & $-1.2(-2)$ \nl
&  &  & & & & \nl
\multicolumn{1}{l}{C$^{18}$O} &  &  & & & &  \nl
J=1$\rightarrow $0 & $1.0(-3)$ & $1.0(-3)$ & $1.1(-3)$ & {$1.1(-3$)} &
 {$1.3(-3)$} & $1.5(-3)$ \nl
 $4 \rightarrow $3 & $1.1(-4)$ & $2.3(-4)$ & $1.1(-3)$ & {$4.6(-3)$} &
 {$1.4(-2)$} & $3.0(-2)$ \nl
 $7 \rightarrow $6 & $9.7(-7)$ & $8.2(-6)$ & $1.2(-4)$ & {$1.3(-3)$} &
 {$8.4(-3)$} & $3.2(-2)$ \nl
 $10 \rightarrow $9 & $8.8(-9)$ & $6.6(-7)$ & $3.1(-6)$ & {$1.4(-4)$} &
 {$2.5(-3)$} & $1.8(-2)$ \nl
 $13 \rightarrow $12 & $-8.7(-11)$ & $1.2(-8)$ & $-2.9(-8)$ &{$-1.7(-5)$} &
 {$3.5(-4)$} & $6.3(-3)$ \nl
 $16 \rightarrow $15 & $-8.7(-11)$ & $-2.6(-11)$ & $7.1(-9)$ & {$-6.0(-6)$} &
 {$-1.9(-5)$} & $1.5(-3)$ \nl
 $19 \rightarrow $18 & $-9.4(-11)$ & $-9.6(-11)$ & $2.6(-10)$ & {$-7.1(-7)$} &
 {$-1.9(-5)$} & $1.5(-4)$ \nl
 $22 \rightarrow $21 & $-6.1(-11)$ & $-6.2(-11)$ & $-7.8(-11)$ & {$-4.8(-8)$} &
 {$-4.0(-6)$} & $-3.1(-5)$ \nl
&  &  & & & & \nl
\multicolumn{1}{l}{$^{13}$CO} &  &  & & & &  \nl
J=1$\rightarrow $0 & $2.5(-3)$ & $2.5(-3)$ & $2.6(-3)$ & {$2.8(-3$)} &
 {$3.5(-3)$} & $4.8(-3)$ \nl
$4 \rightarrow $3 & $5.0(-4)$ & $7.4(-4)$ & $2.4(-3)$ & {$1.0(-2)$} &
 {$3.3(-2)$} & $7.8(-2)$ \nl
$7 \rightarrow $6 & $1.4(-6)$ & $1.2(-5)$ & $2.7(-5)$ & {$3.1(-3)$} &
 {$2.1(-2)$} & $8.7(-2)$ \nl
$10 \rightarrow $9 & $4.3(-8)$ & $1.4(-6)$ & $-4.9(-6)$ & {$3.5(-4)$} &
 {$6.6(-3)$} & $5.0(-2)$ \nl
$13 \rightarrow $12 & $-4.8(-11)$ & $5.6(-8)$ & $-2.4(-6)$ & {$-7.6(-5)$} &
 {$9.0(-4)$} & $1.9(-2)$ \nl
$16 \rightarrow $15 & $-8.7(-11)$ & $2.2(-10)$ & $5.0(-9)$ & {$-2.8(-5)$} &
 {$-1.5(-4)$} & $5.0(-3)$ \nl
$19 \rightarrow $18 & $-9.4(-11)$ & $-9.6(-11)$ & $1.7(-9)$ & {$-3.5(-6)$} &
 {$-9.3(-5)$} & $5.6(-4)$ \nl
$22 \rightarrow $21 & $-6.1(-11)$ & $-6.2(-11)$ & $-4.3(-11)$ & {$-2.4(-7)$} &
 {$-2.0(-5)$} & $-1.5(-4)$ \nl
&  &  &  & &  & \nl
\multicolumn{1}{l}{o-H$_{2}^{16}$O} &  &  & & & &  \nl
$1_{10} \rightarrow 1_{01}$ & $1.0(-3)$ & $9.6(-4)$ & $1.1(-3)$ &
    {$2.3(-3$)} & {$5.2(-3)$} & $3.5(-3)$ \nl
$3_{12} \rightarrow 3_{03}$ & $-5.8(-7)$ & $-5.1(-5)$ & $-5.0(-4)$ &
    {$-1.7(-3)$} & {$1.1(-2)$} & $1.9(-1)$ \nl
$3_{12} \rightarrow 2_{21}$ & $3.0(-7)$ & $-2.3(-5)$ & $-2.4(-4)$ &
    {$-2.1(-4)$} & {$1.5(-2)$} & $1.9(-1)$ \nl
$3_{21} \rightarrow 3_{12}$ & $1.1(-6)$ & $-1.6(-5)$ & $-1.3(-4)$ &
    {$1.7(-3)$} & {$4.0(-2)$} & $4.3(-1)$ \nl
$2_{21} \rightarrow 2_{12}$ & $1.9(-4)$ & $2.4(-5)$ & $-2.1(-3)$ &
    {$-1.5(-2)$} & {$-5.1(-2)$} & $-8.7(-2)$ \nl
$2_{12} \rightarrow 1_{01}$ & $-2.0(-4)$ & $-5.9(-4)$ & $-5.3(-3)$ &
    {$-3.8(-2)$} & {$-1.9(-1)$} & $-7.4(-1)$ \nl
$3_{03} \rightarrow 2_{12}$ & $-6.2(-6)$ & $-2.0(-4)$ & $-2.8(-3)$ &
    {$-2.1(-2)$} & {$-9.6(-2)$} & $-2.1(-1)$ \nl
$2_{21} \rightarrow 1_{10}$ & $-3.1(-4)$ & $-3.6(-4)$ & $-2.5(-3)$ &
    {$-4.2(-2)$} & {$-4.0(-1)$} & $-2.2(-0)$ \nl
$3_{21} \rightarrow 2_{12}$ & $-1.8(-6)$ & $-2.8(-6)$ & $-2.6(-4)$ &
   $-1.6(-2)$ & $-3.4(-1)$ & $-3.5(-0)$ \nl
&  &  &  & &  & \nl
\multicolumn{1}{l}{o-H$_{2}^{18}$O} &  &  & & & &  \nl
$1_{10} \rightarrow 1_{01}$ & $-7.8(-6)$ & $-1.9(-5)$ & $-2.5(-5)$ &
    {$-8.3(-5$)} & {$-9.0(-5)$} & $7.0(-4)$ \nl
$3_{12} \rightarrow 3_{03}$ & $2.5(-8)$ & $5.4(-8)$ & $-4.3(-6)$ &
    {$1.8(-4)$} & {$2.7(-3)$} & $1.8(-2)$ \nl
$3_{12} \rightarrow 2_{21}$ & $1.2(-9)$ & $-4.6(-9)$ & $-8.7(-7)$ &
    {$6.2(-5)$} & {$6.7(-4)$} & $4.6(-3)$ \nl
$3_{21} \rightarrow 3_{12}$ & $5.7(-9)$ & $4.8(-7)$ & $1.2(-5)$ &
    {$3.4(-4)$} & {$4.0(-3)$} & $2.5(-2)$ \nl
$2_{21} \rightarrow 2_{12}$ & $4.3(-8)$ & $-8.1(-6)$ & $-1.3(-4)$ &
    {$-1.0(-3)$} & {$-4.7(-3)$} & $-1.8(-2)$ \nl
$2_{12} \rightarrow 1_{01}$ & $-2.8(-5)$ & $-2.6(-4)$ & $-2.6(-3)$ &
    {$-1.5(-2)$} & {$-5.5(-2)$} & $-1.4(-1)$ \nl
$3_{03} \rightarrow 2_{12}$ & $-2.4(-7)$ & $-1.7(-5)$ & $-3.2(-4)$ &
    {$-2.4(-3)$} & {$-7.6(-3)$} & $-1.0(-2)$ \nl
$2_{21} \rightarrow 1_{10}$ & $-3.9(-7)$ & $-1.8(-5)$ & $-7.3(-4)$ &
    {$-9.6(-3)$} & {$-5.5(-2)$} & $-1.8(-1)$ \nl
$3_{21} \rightarrow 2_{12}$ & $-1.3(-9)$ & $-4.7(-8)$ & $-2.1(-5)$ &
   $-1.5(-3)$ & $-2.8(-2)$ & $-2.0(-1)$ \nl
&  &  &  & &  & \nl
\multicolumn{1}{l}{p-H$_{2}^{16}$O} &  &  & & & &  \nl
$3_{13} \rightarrow 2_{20}$ & $3.0(-8)$ & $7.4(-7)$ & $5.5(-6)$ &
    {$4.0(-5$)} & {$2.5(-4)$} & $7.0(-4)$ \nl
$2_{11} \rightarrow 2_{02}$ & $3.9(-6)$ & $-7.4(-6)$ & $1.1(-4)$ &
    {$2.5(-3)$} & {$2.5(-2)$} & $1.8(-2)$ \nl
$2_{02} \rightarrow 1_{11}$ & $-1.6(-5)$ & $-1.8(-4)$ & $-8.2(-4)$ &
    {$-8.4(-4)$} & {$2.4(-2)$} & $4.6(-3)$ \nl
$1_{11} \rightarrow 0_{00}$ & $-6.0(-4)$ & $-1.1(-3)$ & $-3.8(-3)$ &
    {$-1.5(-2)$} & {$-5.3(-2)$} & $2.5(-2)$ \nl
$2_{20} \rightarrow 2_{11}$ & $7.3(-6)$ & $-5.8(-5)$ & $-5.5(-4)$ &
    {$-1.0(-4)$} & {$3.6(-2)$} & $-1.8(-2)$ \nl
$3_{22} \rightarrow 3_{13}$ & $-3.2(-8)$ & $-1.4(-5)$ & $-5.1(-4)$ &
    {$-6.0(-3)$} & {$-2.9(-2)$} & $-1.4(-1)$ \nl
$3_{13} \rightarrow 2_{02}$ & $-1.1(-6)$ & $-8.8(-5)$ & $-2.3(-3)$ &
    {$-2.6(-2)$} & {$-1.6(-1)$} & $-1.0(-2)$ \nl
$2_{20} \rightarrow 1_{11}$ & $-2.2(-5)$ & $-4.6(-5)$ & $-1.6(-3)$ &
    {$-3.6(-2)$} & {$-4.0(-1)$} & $-1.8(-1)$ \nl
$3_{22} \rightarrow 2_{11}$ & $-2.9(-9)$ & $-3.1(-6)$ & $-4.8(-4)$ &
   $-1.8(-2)$ & $-2.6(-1)$ & $-2.0(-1)$ \nl
&  &  &  & &  & \nl
\multicolumn{1}{l}{p-H$_{2}^{18}$O} &  &  & & & &  \nl
$3_{13} \rightarrow 2_{20}$ & $-5.6(-10)$ & $1.8(-9)$ & $2.2(-8)$ &
    {$5.1(-7$)} & {$1.0(-5)$} & $7.0(-4)$ \nl
$2_{11} \rightarrow 2_{02}$ & $6.7(-7)$ & $5.7(-6)$ & $6.2(-5)$ &
    {$6.0(-4)$} & {$3.1(-3)$} & $1.8(-2)$ \nl
$2_{02} \rightarrow 1_{11}$ & $-5.9(-7)$ & $-9.5(-6)$ & $-4.4(-6)$ &
    {$5.8(-4)$} & {$4.7(-3)$} & $4.6(-3)$ \nl
$1_{11} \rightarrow 0_{00}$ & $-4.6(-5)$ & $-2.6(-4)$ & $-1.3(-3)$ &
    {$-4.1(-3)$} & {$-9.3(-3)$} & $2.5(-2)$ \nl
$2_{20} \rightarrow 2_{11}$ & $2.4(-8)$ & $5.8(-7)$ & $3.6(-5)$ &
    {$6.9(-4)$} & {$4.8(-3)$} & $-1.8(-2)$ \nl
$3_{22} \rightarrow 3_{13}$ & $-1.9(-10)$ & $-7.1(-8)$ & $-4.9(-6)$ &
    {$-3.5(-5)$} & {$-3.5(-6)$} & $-1.4(-1)$ \nl
$3_{13} \rightarrow 2_{02}$ & $-2.5(-8)$ & $-4.3(-6)$ & $-1.5(-4)$ &
    {$-1.6(-3)$} & {$-7.9(-3)$} & $-1.0(-2)$ \nl
$2_{20} \rightarrow 1_{11}$ & $-2.5(-8)$ & $-4.1(-6)$ & $-2.8(-4)$ &
    {$-5.2(-3)$} & {$-3.8(-2)$} & $-1.8(-1)$ \nl
$3_{22} \rightarrow 2_{11}$ & $-9.0(-11)$ & $-3.5(-8)$ & $-9.8(-6)$ &
   $-4.9(-4)$ & $-6.8(-3)$ & $-2.0(-1)$ \nl
&  &  &  & &  & \nl
\multicolumn{1}{l}{${}^{16}$O} &  &  & & & &  \nl
$0 \rightarrow 1$ & $-3.1(-9)$ & $-1.2(-7)$ & $-2.3(-5)$ &
    $-4.3(-4)$ & $1.7(-3)$ & $8.6(-2)$ \nl
$1 \rightarrow 2$ & $-2.4(-6)$ & $-2.5(-6)$ & $-5.7(-5)$ &
    $-6.0(-3)$ & $-1.8(-1)$ & $-2.0(-0)$ \nl
&  &  &  & &  & \nl
\enddata
\tablecomments{line luminosities for the subset of models with
$M_{0.1}=100$ M$_{\odot}$.  
In this table, $a(b)$ means $a {\times} 10^{b}$ L$_\odot$.
See Table 6 
 for rest frequencies.}
\label{tbl5}
\end{deluxetable}

\clearpage

\begin{table}
\begin{center}
\caption{
Maximum upper
rotational quantum number, $J_u$, for any $^{12}$CO
emission line.
}
\vspace{0.3cm}
\begin{tabular}{lccccccc} \hline \hline
$M_{0.1}/$M$_{\odot}$ &
$1$ L$_{\odot}$ &
$10$ L$_{\odot}$ &
$10^{2}$ L$_{\odot}$ &
$10^{3}$ L$_{\odot}$ &
$10^{4}$ L$_{\odot}$ &
$10^{5}$ L$_{\odot}$ &
$10^{6}$ L$_{\odot}$
\\
\hline

$1$ & 24 & 24 & 24 & --- & --- & --- & --- \\
$10$ & 22 & 24 & 24 & 24 & --- & --- & --- \\
$10^2$ & 13$^a$ & 18$^b$ & 24$^c$ & 11 & 14 & 17 & --- \\
$10^3$ & 8 & 5 & 5 & 7 & 9 & 11 & 12 \\
\hline
\hline
\end{tabular}
\end{center}
\label{tbl7}
$^a$ Transitions with $J_u=7$ and 8 are absorption lines\\
$^b$ Transitions with $8 \leq J_u \leq 12$ are absorption lines\\
$^c$ Transitions with $9 \leq J_u \leq 20$ are absorption lines

\end{table}

\clearpage

\begin{table}
\begin{center}
\caption{
Target water lines for high-z extra-galactic sources
}
\vspace{0.3cm}
\begin{tabular}{rrrrr} \hline \hline
{} &
{Rest} &
{Redshifted} &
{Frequency} &
{} \\
{} &
{Frequency} &
{F10214+4724} &
{Cloverleaf} &
{BR1202-0725} \\
{Transition} &
{(GHz)} &
{z=2.286} &
{z=2.551} &
{z=4.702}

\\ \hline

$2_{11} \rightarrow 2_{02}$ & $752.0$ & $228.9$ & $211.8$ & $131.9$ \nl
$2_{02} \rightarrow 1_{11}$ & $987.9$ &         &         & $173.2$ \nl
$3_{12} \rightarrow 3_{03}$ & $1097$  &         &         & $192.4$ \nl
$1_{11} \rightarrow 0_{00}$ & $1113$  &         &         & $195.3$ \nl
$3_{12} \rightarrow 2_{21}$ & $1153$  &         &         & $202.2$ \nl
$3_{21} \rightarrow 3_{12}$ & $1163$  &         &         & $203.9$ \nl
$2_{20} \rightarrow 2_{11}$ & $1229$  &         &         & $215.5$ \nl
\\
\hline
\hline
\end{tabular}
\end{center}
\label{tbl9}
\end{table}

\clearpage

\begin{table}
\begin{center}
\caption{
Effective $^{12}$CO and H$_{2}^{16}$O
column densities, for the $v=2-0$ and $\nu_2$ absorption bands
respectively.
In this table $a(b)$ means $a \times 10^{b}\,\rm cm^{-2}$.
}
\vspace{0.3cm}
\begin{tabular}{lrccccccc} \hline \hline
Species  &
$M_{0.1}/M_{\odot}$ &
$1$ L$_{\odot}$ &
$10$ L$_{\odot}$ &
$10^{2}$ L$_{\odot}$ &
$10^{3}$ L$_{\odot}$ &
$10^{4}$ L$_{\odot}$ &
$10^{5}$ L$_{\odot}$ &
$10^{6}$ L$_{\odot}$
\\
\hline
CO & 1   & 1.0(19) & 3.4(19) & 4.9(19) &  --- & --- & --- & --- \\
   & 10  & 1.4(18) & 1.3(18) & 2.2(18) & 1.2(20) & --- & --- & --- \\ 
   & 60  & ---  & ---  & ---  & --- & --- & 1.1(20) & --- \\
   & $10^2$ & 2.9(18) & 2.9(18) & 2.9(18) & 2.8(18) & 1.1(19) & 1.5(20)
               & --- \\ 
   & $10^3$ & 5.7(18) & 5.7(18) & 5.7(18) & 5.7(18) & 5.6(18) & 5.4(18)
               & 5.3(19) \\
 & & & & & & & & \nl
H$_{2}$O & 1 & 5.0(16) & 1.5(17) & 5.9(18) & --- & --- & --- & ---  \\
   & 10 & 3.8(16) & 3.7(16) & 3.0(17) & 9.2(18) & --- & --- & --- \\
   & 60 & --- & --- & --- & --- & --- & 1.4(19) & --- \\
   & $10^{2}$ & 9.4(16) & 9.4(16) & 9.4(16) & 9.1(16) & 1.5(18) & 8.1(18) 
          & --- \\
   & $10^{3}$ & 2.1(17) & 2.1(17) & 2.1(17) & 2.1(17) & 2.1(17) & 2.0(17)
          & 1.6(18) \\ 
\hline
\hline
\end{tabular}
\end{center}
\label{tbl8}
\end{table}

\clearpage 

\figcaption{Schematic representation of the model presented in this paper.}
 
\figcaption{Comparison of the predicted dust continuum spectrum for the
standard
model (solid line) with observations of the Orion
Molecular Cloud 1 (symbols).}

\figcaption{Dust (dashed line) and gas (solid line) temperature distributions
for the standard model.}

\figcaption{Fractional abundances of $^{12}$C$^{16}$O, atomic oxygen, and
o-H$_{2}^{16}$O relative to H$_2$, as functions of position in our
standard model.}

\figcaption{Cooling coefficient ($\Lambda/n[\rm H_2]^2$) for a variety of
species in our standard model.} 

\figcaption{Line luminosities for the three isotopes of CO, as a function of
upper rotational quantum number for our standard model.}

\figcaption{Continuum and line ($^{12}$CO, H$_{2}^{16}$O, and OI) 
spectrum for the standard model, assuming D = 1 kpc and a resolution
of $\lambda / \Delta \lambda$ = 10$^4$.}
 
\figcaption{Line profiles for the nine transitions of ortho-H$_2^{16}$O
included in our standard model.  The vertical axis shows the monochromatic
luminosity.}

\figcaption{Same as Figure 8, but for transitions of ortho-H$_2^{18}$O.}

\figcaption{Same as Figure 8, but for transitions of para-H$_2^{16}$O and
atomic oxygen.}

\figcaption{Same as Figure 8, but for transitions of para-H$_2^{18}$O.} 

\figcaption{Gas (solid lines) and dust (dashed lines) temperature
distributions for all models in our grid.  Each box corresponds to
a given value of $M_{0.1}$, and each curve is labeled by 
${\rm log}_{10}(L_{*})$.}  

\figcaption{Line strengths for selected $^{12}$CO transitions for the
grid of models considered in the text. The upper panel of each box
corresponds to emission for a given transition, while the lower panel of the
same box corresponds to absorption for the same transition. The different
symbols correspond to different core masses ($M_{0.1}$): the open
triangles, open squares, crosses, and open circles correspond to $M_{0.1}=1,$
$10,$ $10^2,$ and $10^3$ M$_{\odot }$ respectively.
Each box is labeled with the molecular transition to which it refers.}

\figcaption{Same as Figure 13, but for transitions of ortho-H$_2^{16}$O.}

\figcaption{Same as Figure 13, but for transitions of ortho-H$_2^{18}$O.}

\figcaption{Same as Figure 13, but for transitions of para-H$_2^{16}$O.}

\figcaption{Same as Figure 13, but for transitions of para-H$_2^{18}$O
and atomic oxygen.}

\figcaption{$^{12}$CO emission line luminosity as a function
of upper rotational quantum number, $J_u$, for various value of the 
core mass, $M_{0.1}$.  Results apply to cloud cores in which the
central luminosity $L_* = 100\,\rm L_\odot$.}


\clearpage




\begin{references}
\reference{}Abgrall, H. Le Bourlot, J., Pineau des Forets, G., Roueff, E.,  
               Flower, D. R., \& Heck, L. 1992, A\&A, 253, 525 
\reference{}{{Barvainis, R., Antonucci, R., \& Coleman, P. 1992, ApJ, 399, L19}}
\reference{}{{Barvainis, R., Tacconi, L., Antonucci, R., Alloin, D., \&
Coleman, P. 1994, Nature, 371, 586}}
\reference{}Bohlin, R. C., Savage, B. D., \& Drake, J. F. 1978,
ApJ, 224, 132
\reference{}{{Brown, R. L., \& Vanden Bout, P. A. 1991, AJ, 102, 1956}}
\reference{}{{Carr, J. S., Evans, N. J. II, Lacy, J. H., \& Zhou, S. 1995,
ApJ, 450, 667}}
\reference{}Ceccarelli, C., Hollenbach, D. J., \& Tielens, A. G. G. M.
1996, ApJ, 471, 400 (CHT)
\reference{}Cernicharo, J., Gonzalez-Alfonso, E., Alcolea, J., Bachiller, R., \& John D.
               1994, ApJ, 432, L59
\reference{}Cernicharo, J., Thum, C., Hein, H., John, D., Garcia, P., \& Mattioco, F.
               1990, A\&A, 231, L15
\reference{}Doty, S. D. 1997, in preparation
\reference{}{Doty, S. D., \& Neufeld, D. A. 1997, in preparation}
\reference{}{{Downes, D., Solomon, P. M., \& Radford, S. E. J. 1993, ApJ, 414, L13}}
\reference{}Draine, B. T., \& Lee, H. M. 1987, ApJ, 318, 485
\reference{}Egan, M. P., Leung, C. M., \& Spagna, G. F., Jr. 1988,
Comput. Phys. Comm., 48, 271
\reference{}{{Erickson, E. F. et al. 1977, ApJ, 212, 696}} 
\reference{}{{Fuller, G. A., \& Myers, P. C. 1993, ApJ, 418, 273}}
\reference{}{{Gensheimer, P., Mauersberger, R., \& Wilson, T. L. 1996, 
A\&A, in press}}
\reference{}{{Genzel, R., \& Stutzki, J. 1989, ARAA, 27, 41}}
\reference{}{{Goldsmith, P. F., \& Langer, W. D. 1978, ApJ, 222, 881}}
\reference{}{{Goldsmith, P. F., Langer, W. D., Schloerb, F. P., \&
Scoville, N. Z. 1980, ApJ, 240, 524}}
\reference{}{{Green, S., Maluendes, S., \& McLean, A. D. 1993, ApJS, 85, 181}}
\reference{}{{Helmich, F. P., van Dishoeck, E. F., Black, J. H., \&
de Graauw, Th. 1996, A\&A, in press}}
\reference{}{{Hewitt, A., Burbidge, G. 1989 in ``A New Optical Catalog of
Quasi-Stellar Objects"}}
\reference{}{{Hollenbach, D. J., \& McKee, C. F. 1989, ApJ, 342, 306}}
\reference{}{{Jacq, T., Jewell, P. R., Henkel, C., Walmsley, C. M., \& Baudry,
A. 1988, A\&A, 199, L5}}
\reference{}{{Keto, E. R., Ho, P. T. P., \& Haschick, A. D. 1988, ApJ, 324, 920}}
\reference{}{{Larson, R. B. 1969, MNRAS, 145, 271}}
\reference{}{{Mathis, J. S., Mezger, P. G., \& Pangia, N.1983, A\&A, 128, 212}}
\reference{}{{Millar, T. J., Rawlings, J. M. C., Bennett, A., Brown, P. D., \& 
              Charnley, S. B. 1991, A\&AS, 87, 585}}
\reference{}{{Mitchell, G. F., Curry, C., Maillard, J.-P., Allen, M., 1989,
ApJ, 341, 1020}}
\reference{}{{Mitchell, G. F., Maillard, J.-P., Allen, M., Beer, R., 
Belcourt, K., 1990 ApJ, 363, 554}}
\reference{}{{Mozurkewich, D., Schwartz, P. R., \& Smith, H. A.
 1986, ApJ, 311, 371}}
\reference{}{{Myers, P. C., \& Fuller, G. A. 1992, ApJ, 396, 631}}
\reference{}{{Neufeld, D. A., \& Green, S. 1994, ApJ, 432, 158}}
\reference{}{{Neufeld, D. A., \& Kaufman, M. J. 1993, ApJ, 418, 263 (NK)}}
\reference{}{{Neufeld, D. A., Lepp, S., \& Melnick, G. J. 1995, ApJS, 100, 132 (NLM)}}
\reference{}{{Petitjean, P., Pecontal, E., Valls-Gabaud, D., \& Charlot, S. 1996,
Nature, 380, 411}}
\reference{}{{Phillips, T. R. \& Green, S. 1995, Ap\&SS, 224, 537}}
\reference{}{{Phillips, T. G., Kwan, J., \& Huggins P. J. 1980, in
``Interstellar Molecules", ed. B.H. Andrew, Reidel, Dordrecht, Holland, p21.}}
\reference{}{{Rybicki, G. B., \& Hummer, D. G. 1992, A\&A, 262, 209}}
\reference{}{{Shu, F. H. 1977, ApJ, 214, 488}}
\reference{}{{Shu, F. H., Adams, F. C., \& Lizano, S. 1987, ARAA, 25, 23}}
\reference{}{{Takahasi, T., Hollenbach, D. J., \& Silk, J.
 1983, ApJ, 275, 145}}
\reference{}{{Tauber, J., Olofsson, G., Pilbratt, G., Nordh, L., \&
Frisk, U. 1996, A\&A, 308, 913}}
\reference{}{{van Dishoeck, E. F., \& Helmich, F. P. 1996, A\&A, in press 
(vDH)}}
\reference{}{{Wannier, P. G., Pagani, L., Kuiper, T. B. H., Frerking, M. A.,
Gulkis, S., Encrenaz, P., Pickett, H. M., Lecacheux, A., \& Wilson, 
W. J. 1991, ApJ, 377, 171}}
\reference{}{{Waters, J. W., Gustincic, J. J., Kakar, R. K., et al.
1980, ApJ, 235, 57}}
\reference{}{{Whittet, D. C. B. 1992, ``Dust in the Galactic Environment",
IOP Publishing, Ltd.}}
\reference{}{{Zmuidzinas, J., Blake, G. A., Carlstrom, J., Keene, J., 
Miller, D., \& Schilke, P. 1996, ApJ, submitted}}
\end{references}
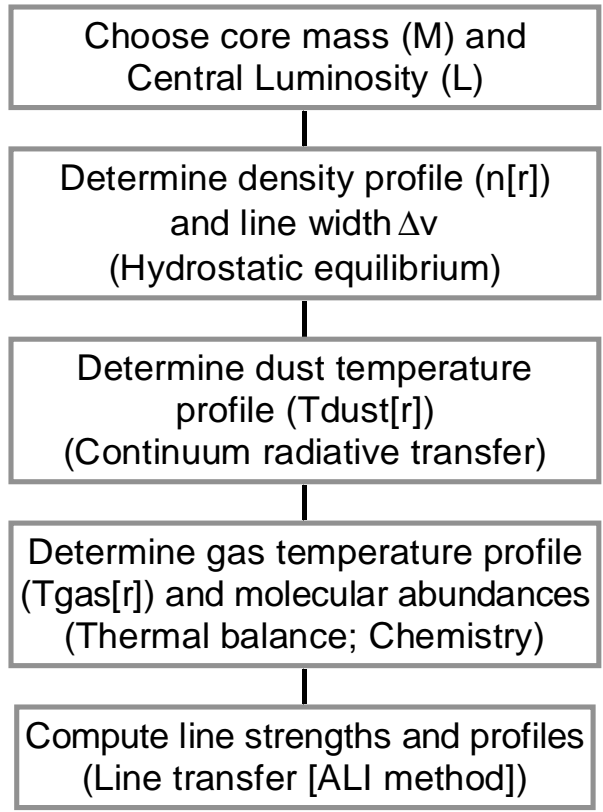
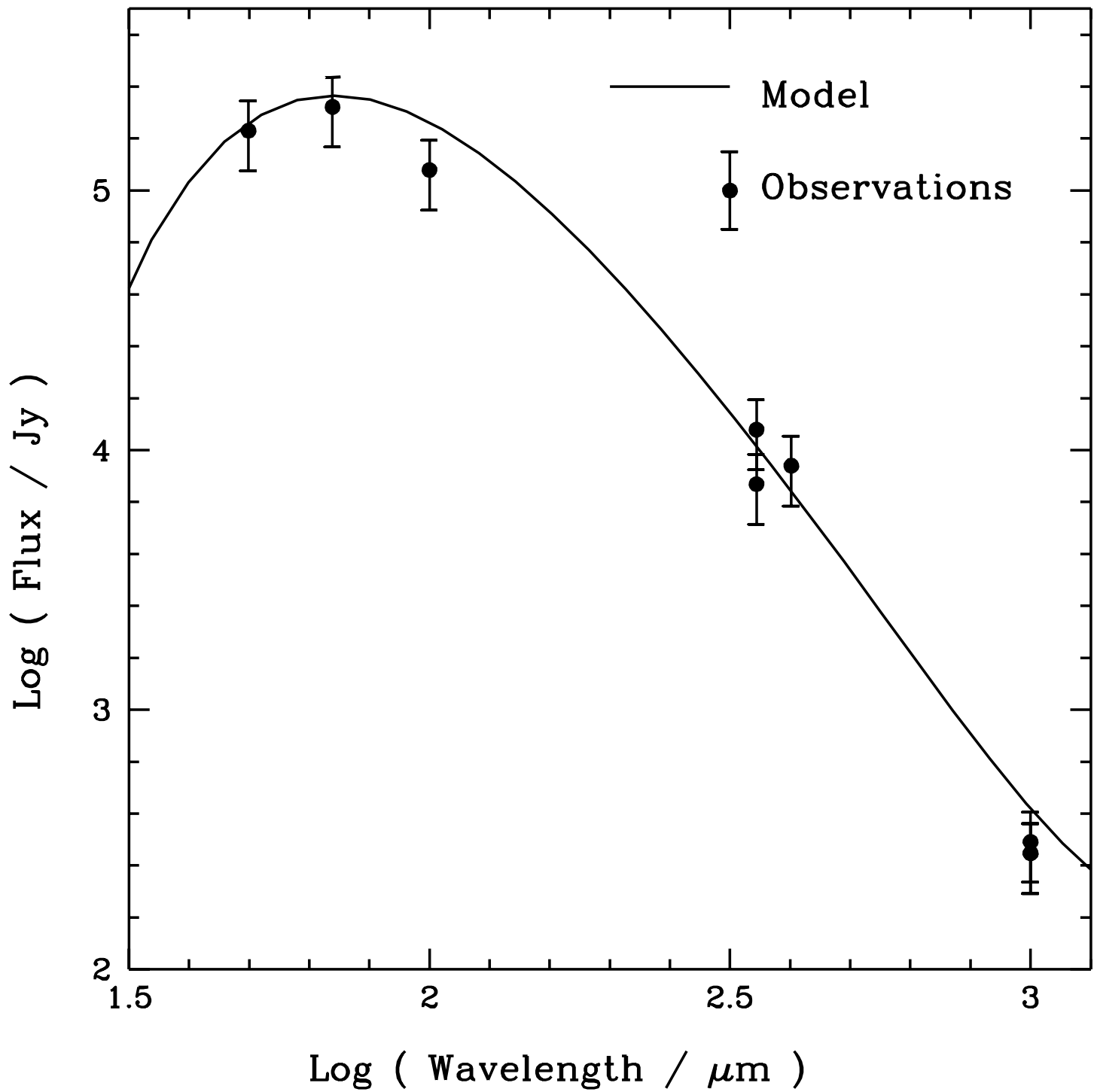
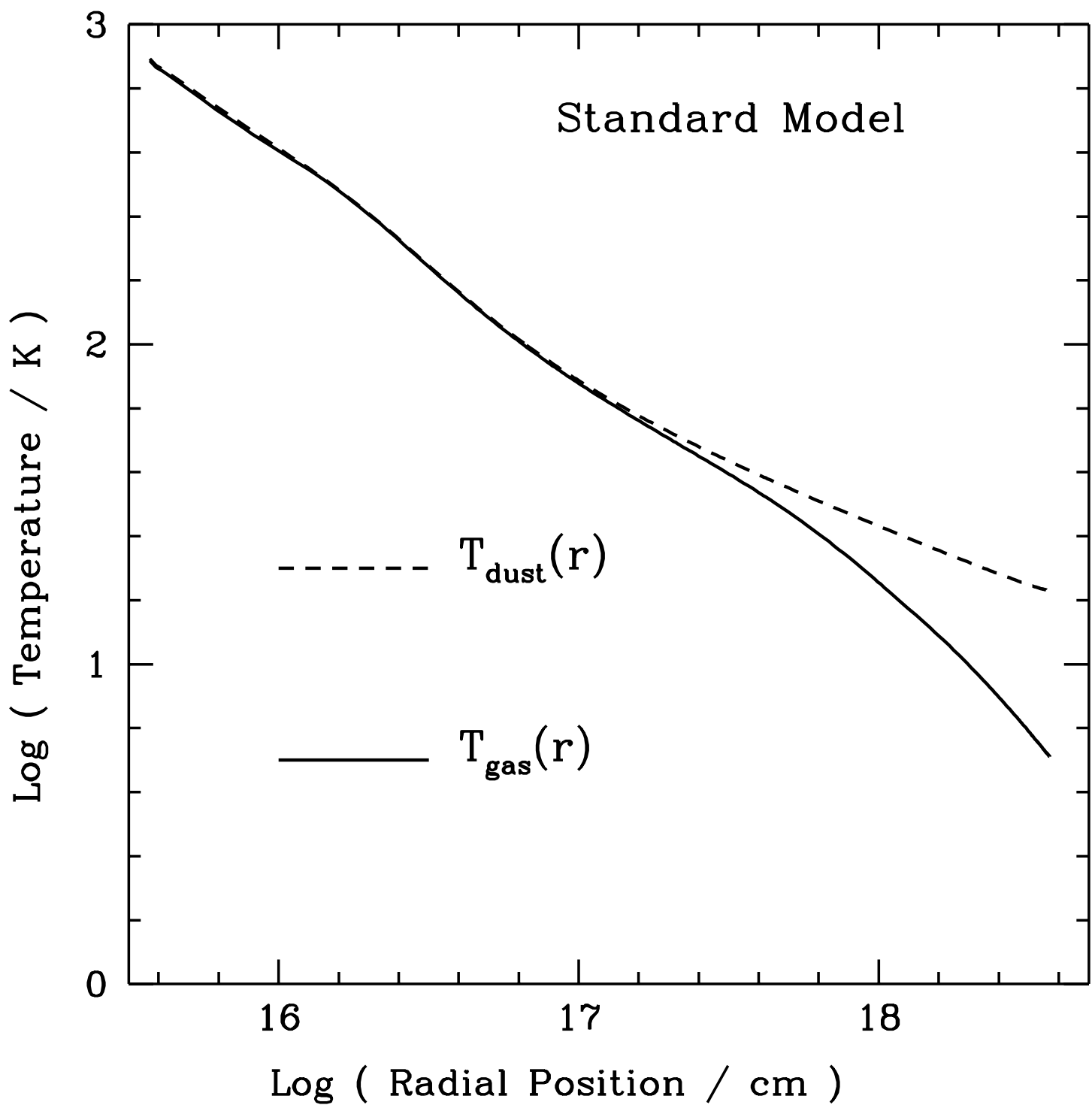
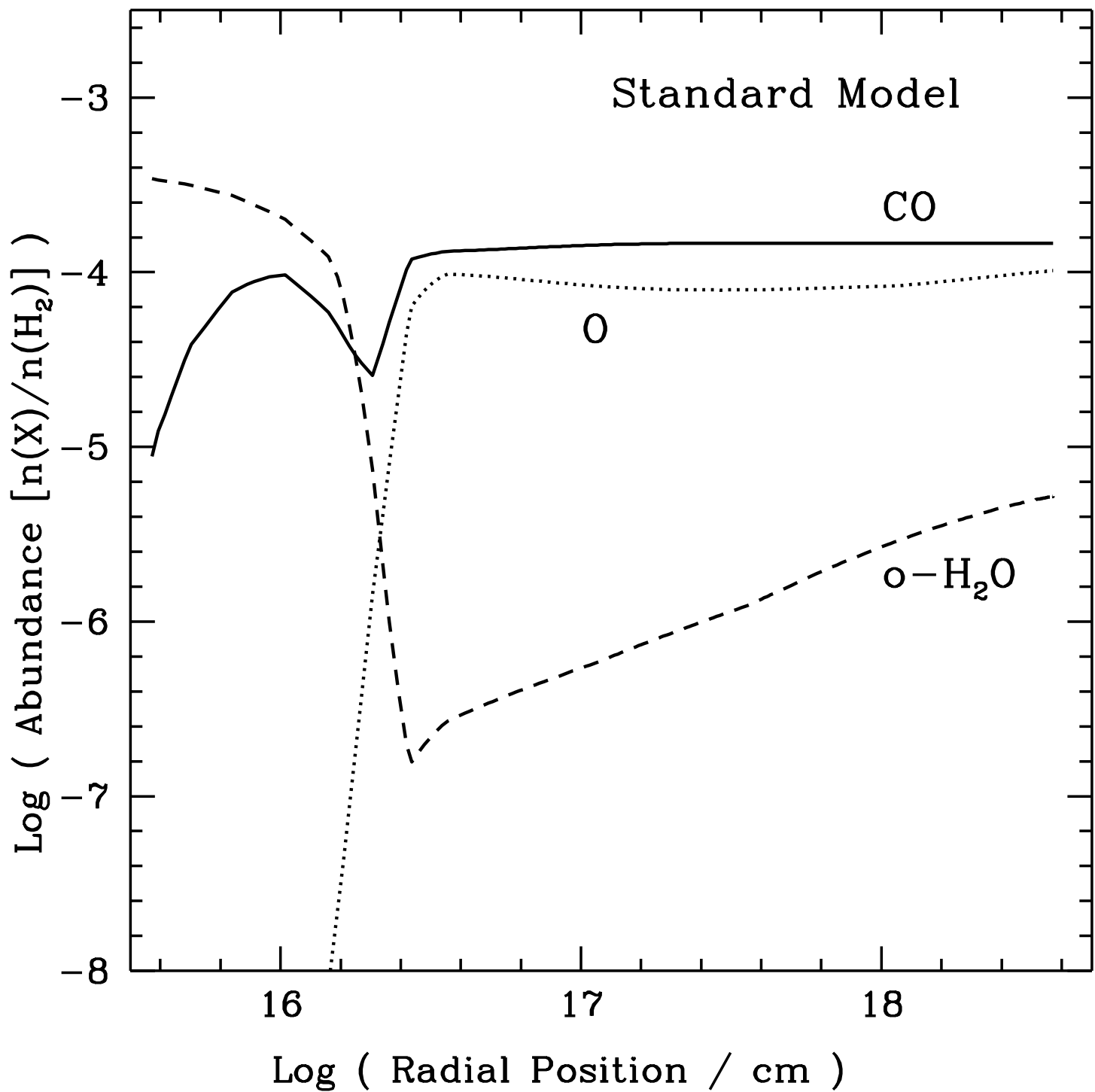
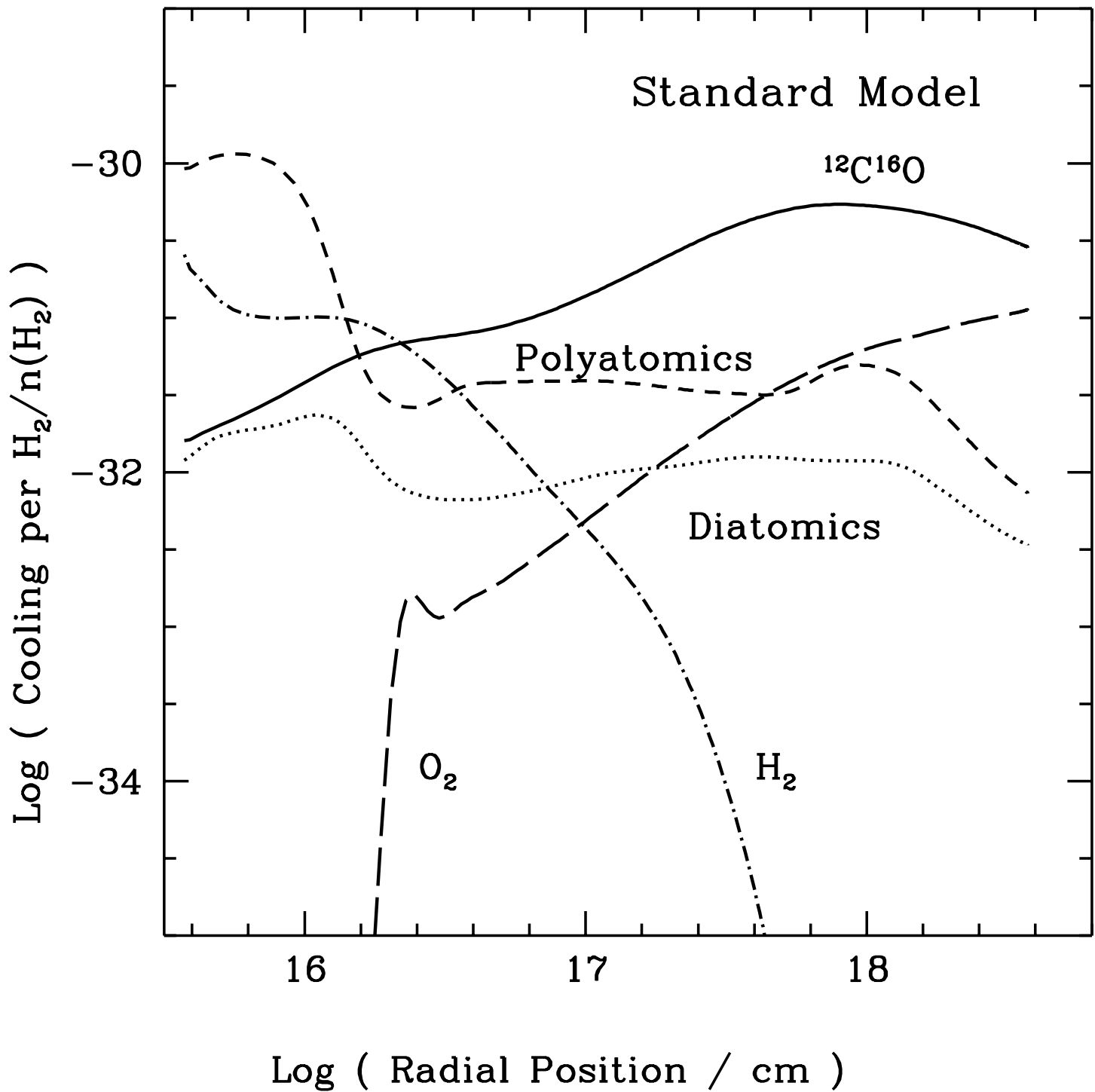
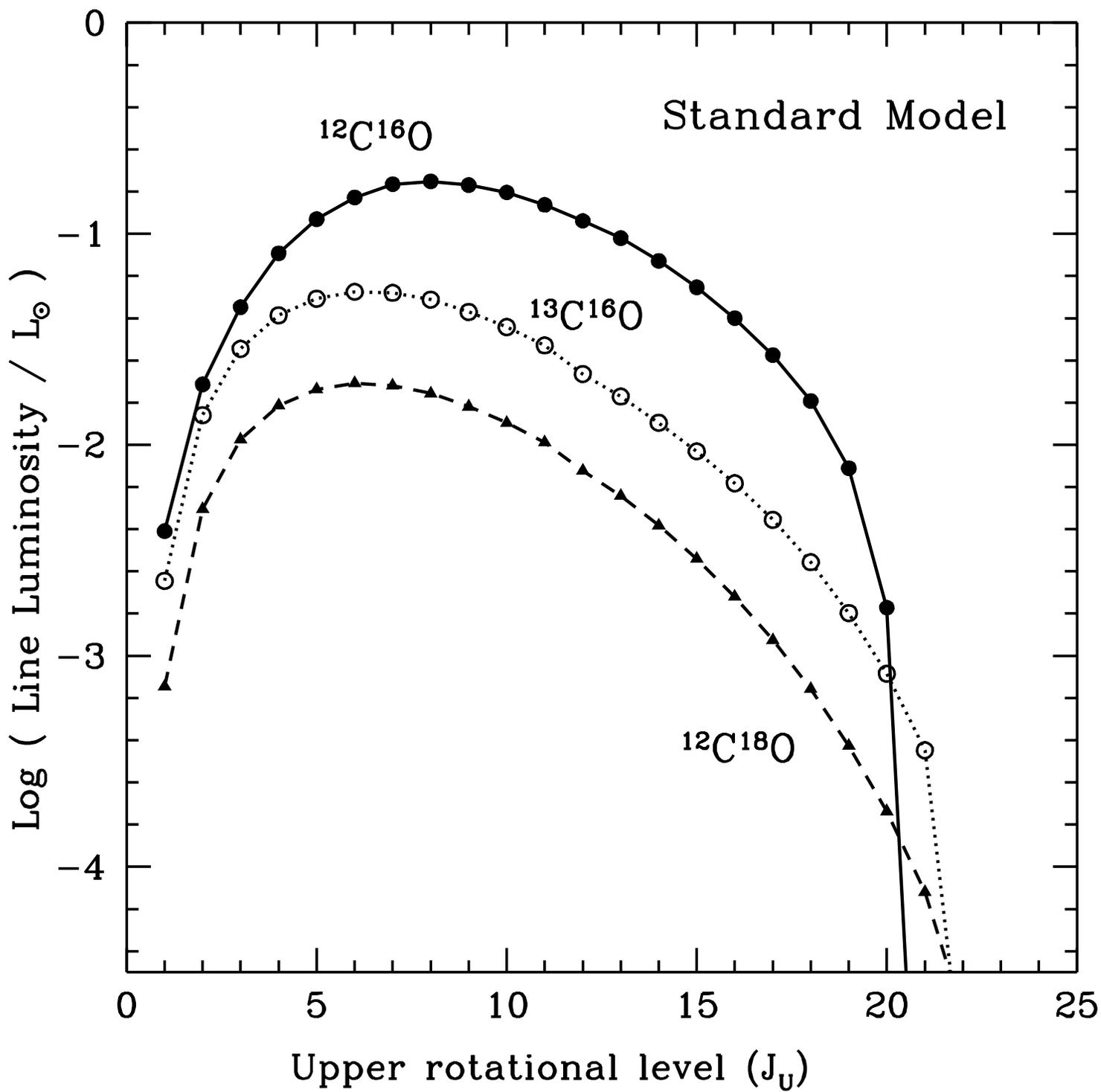
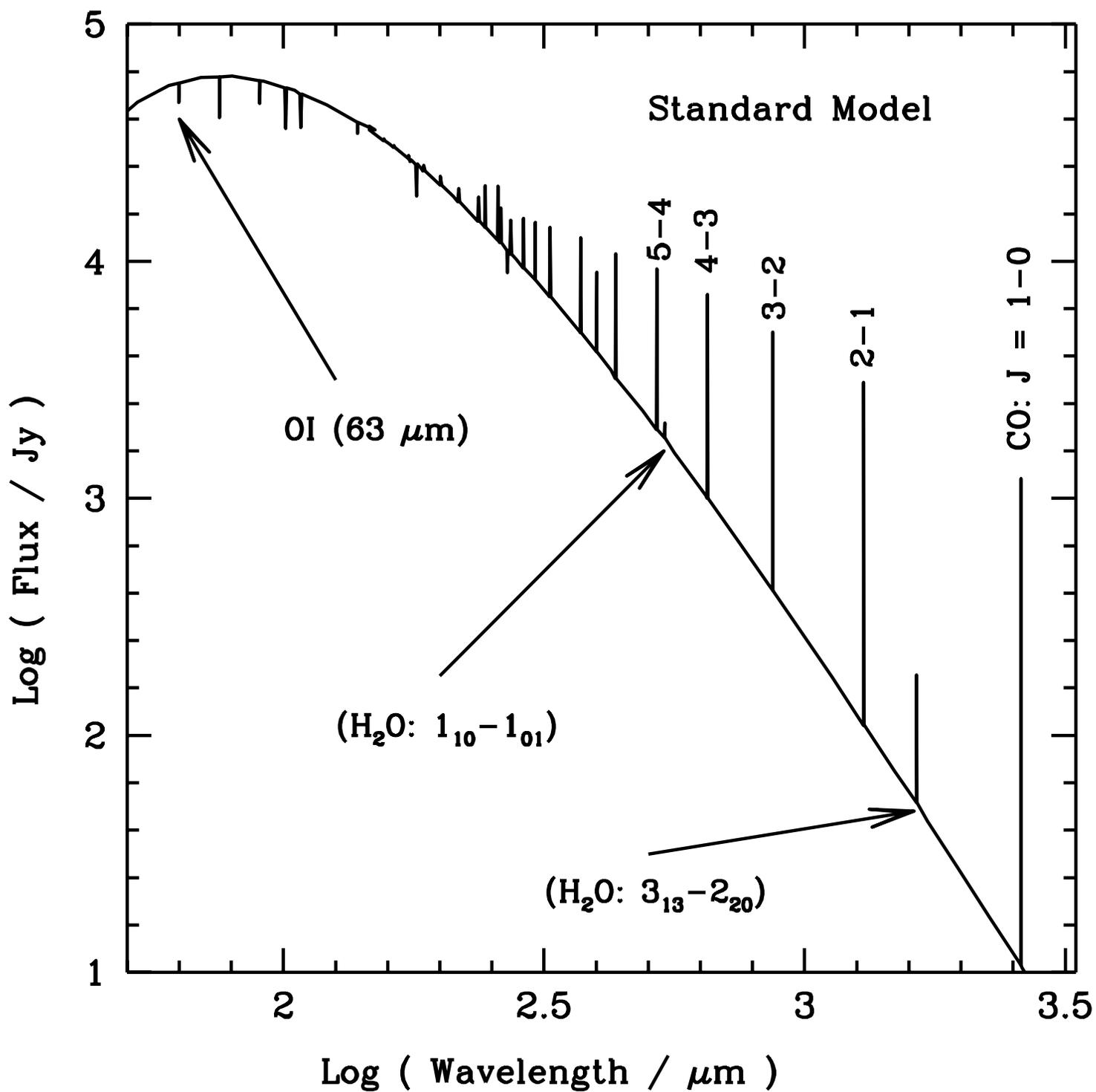
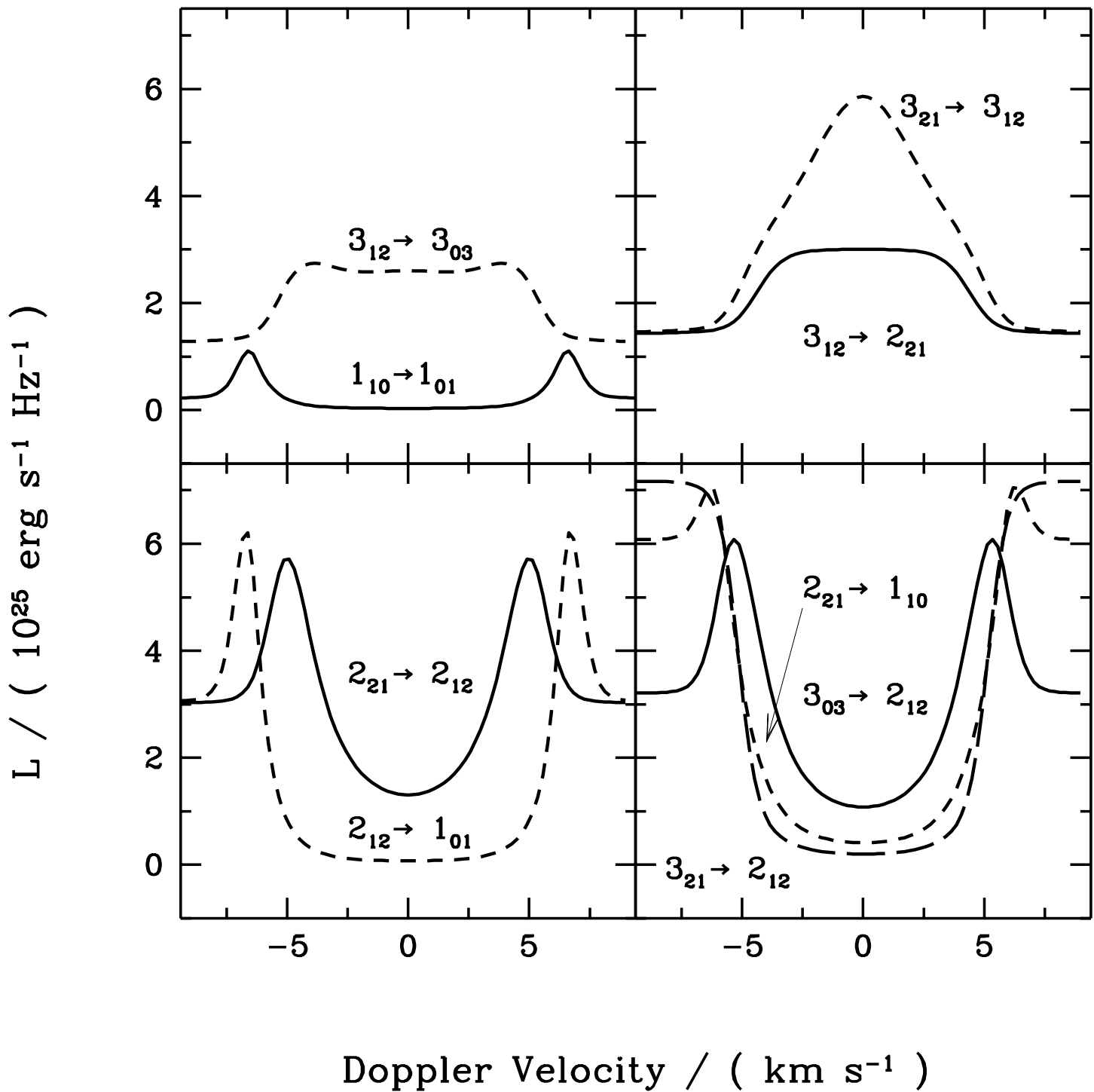
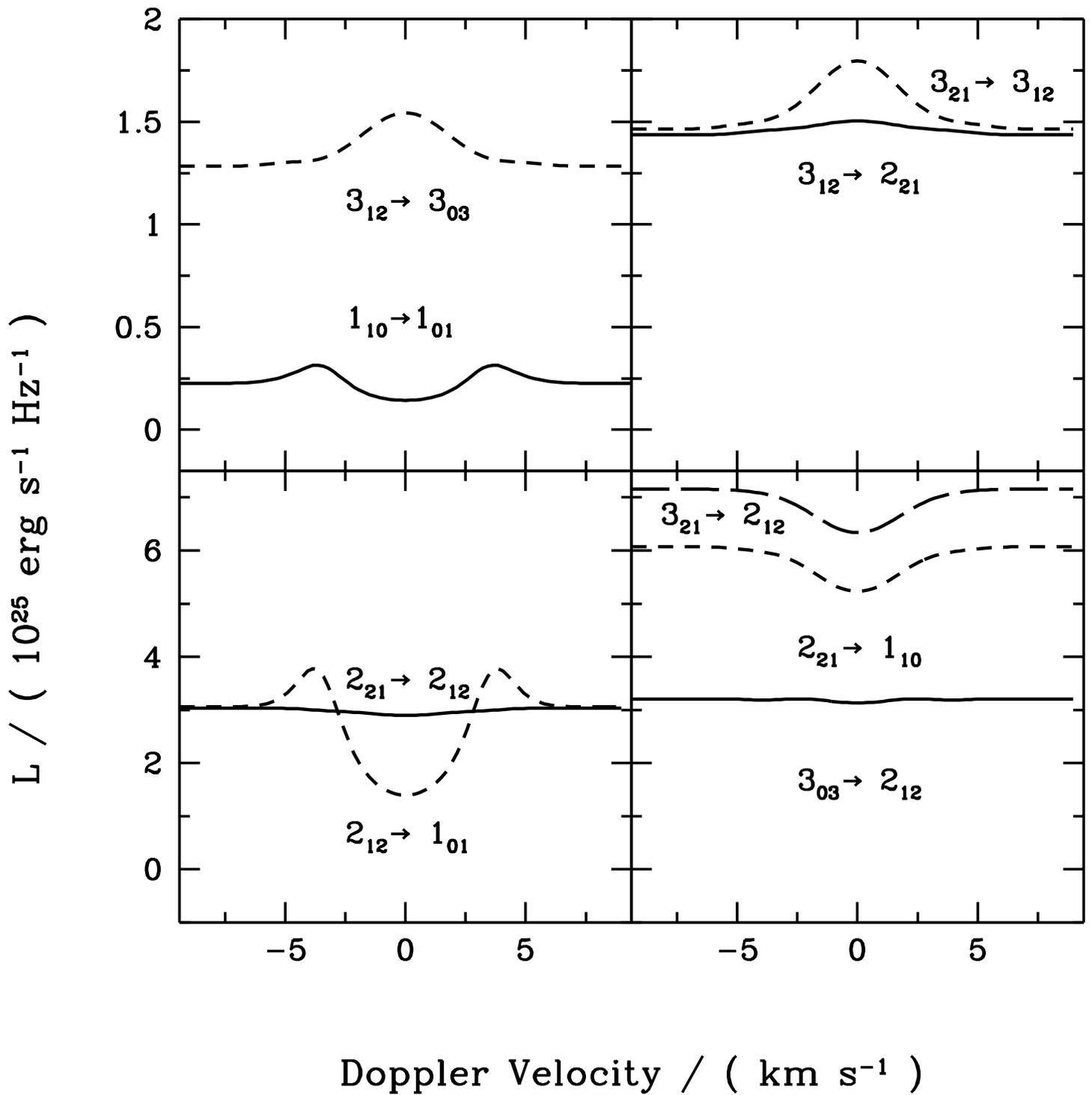
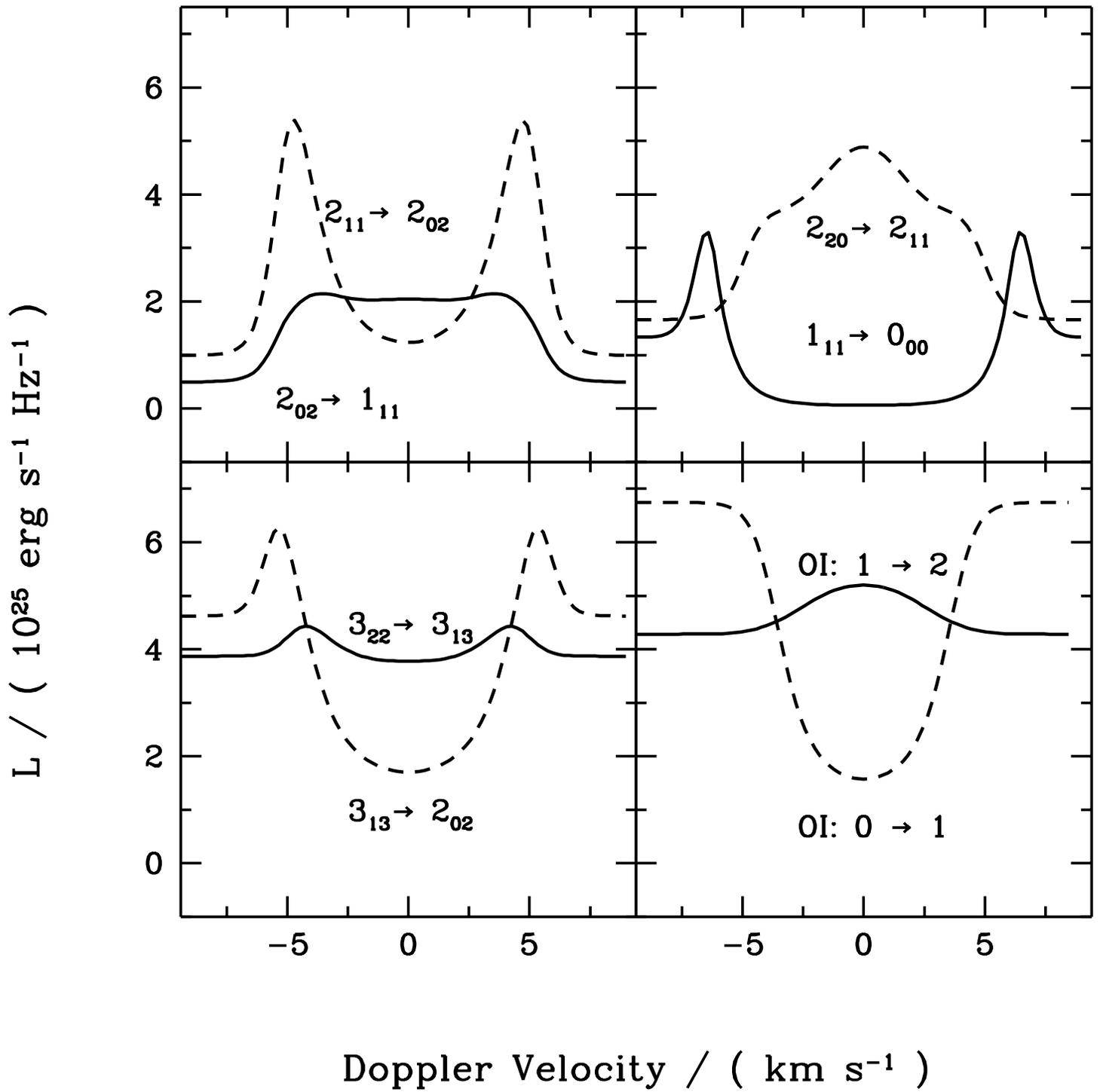
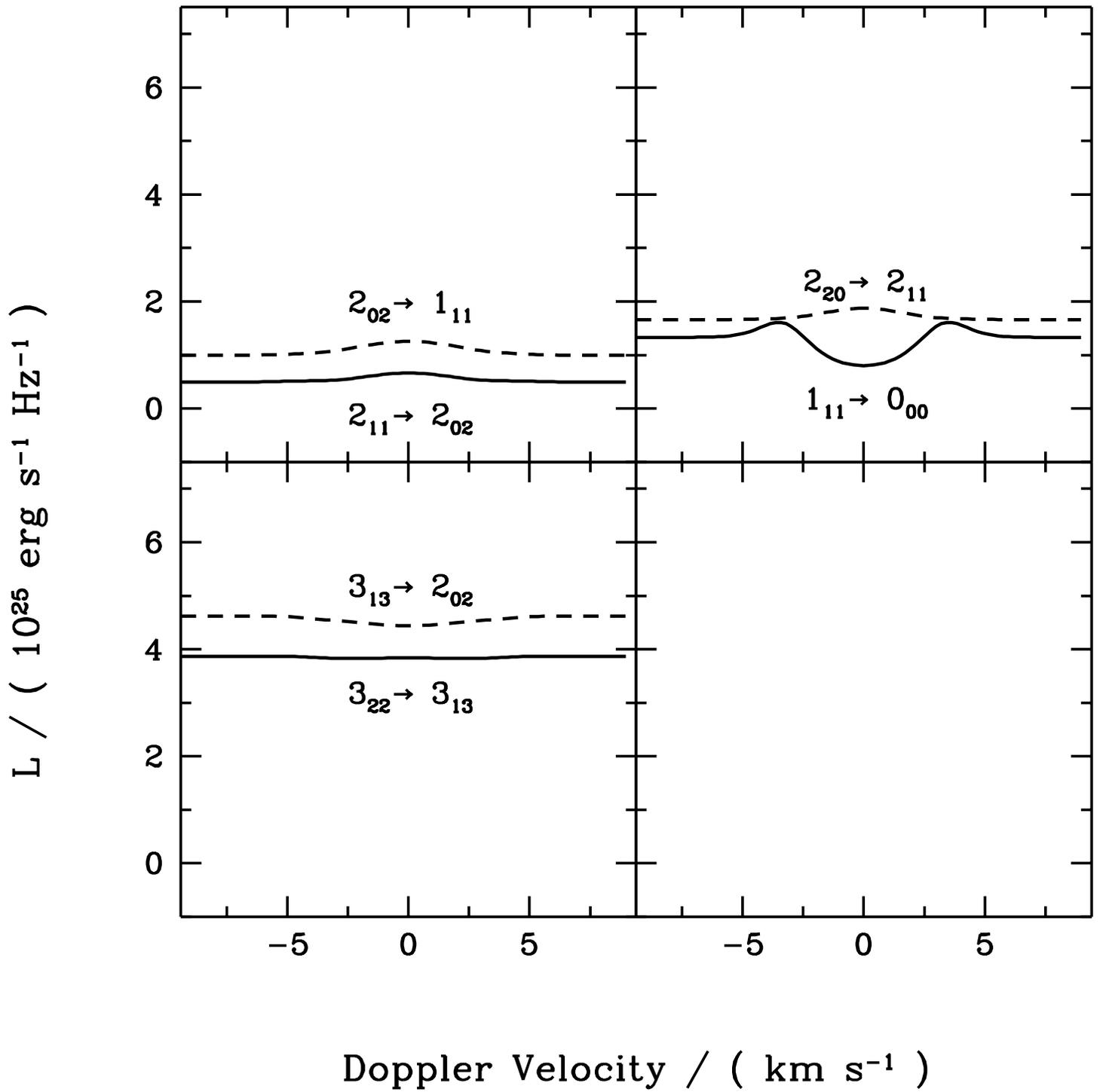
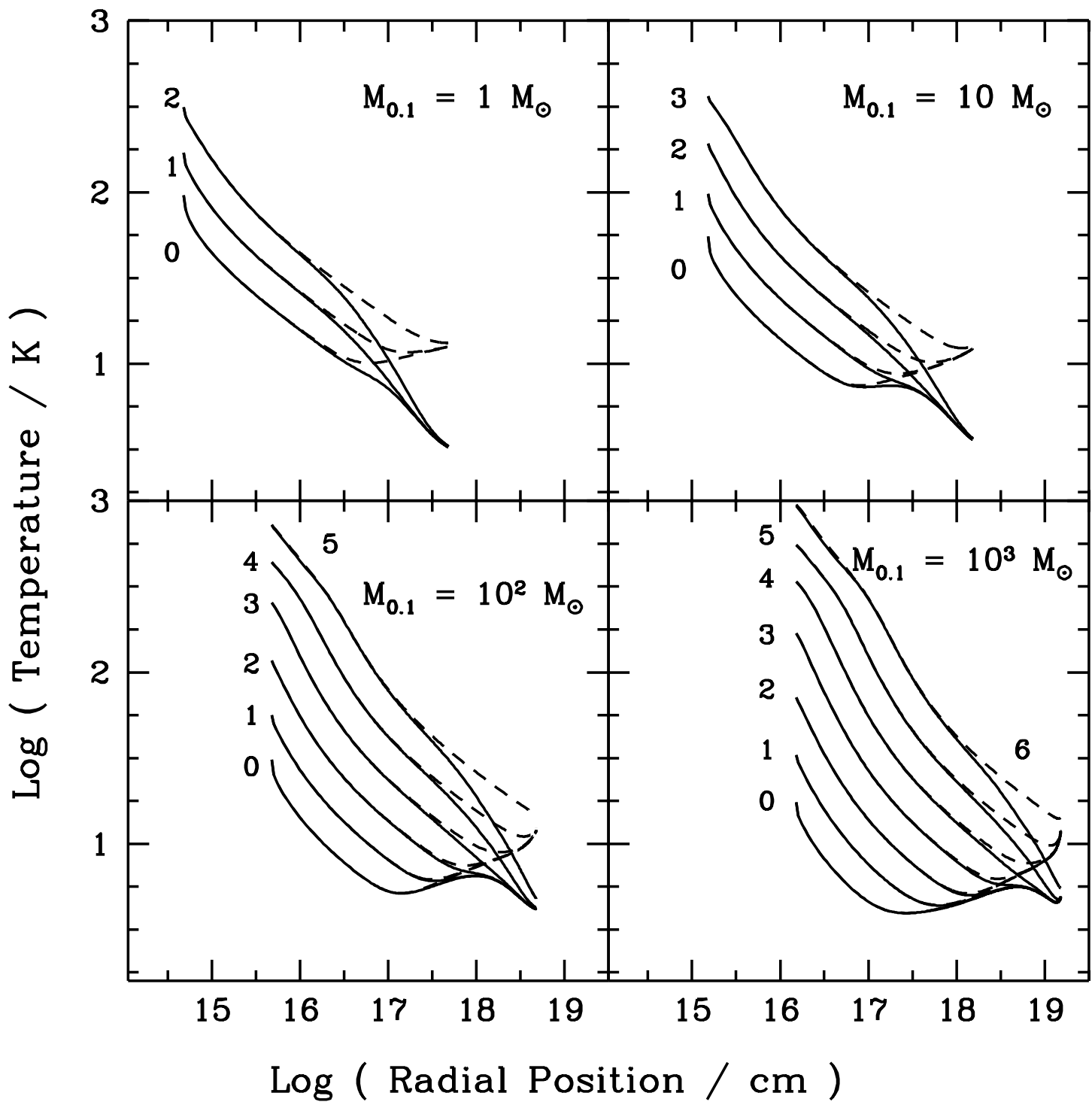
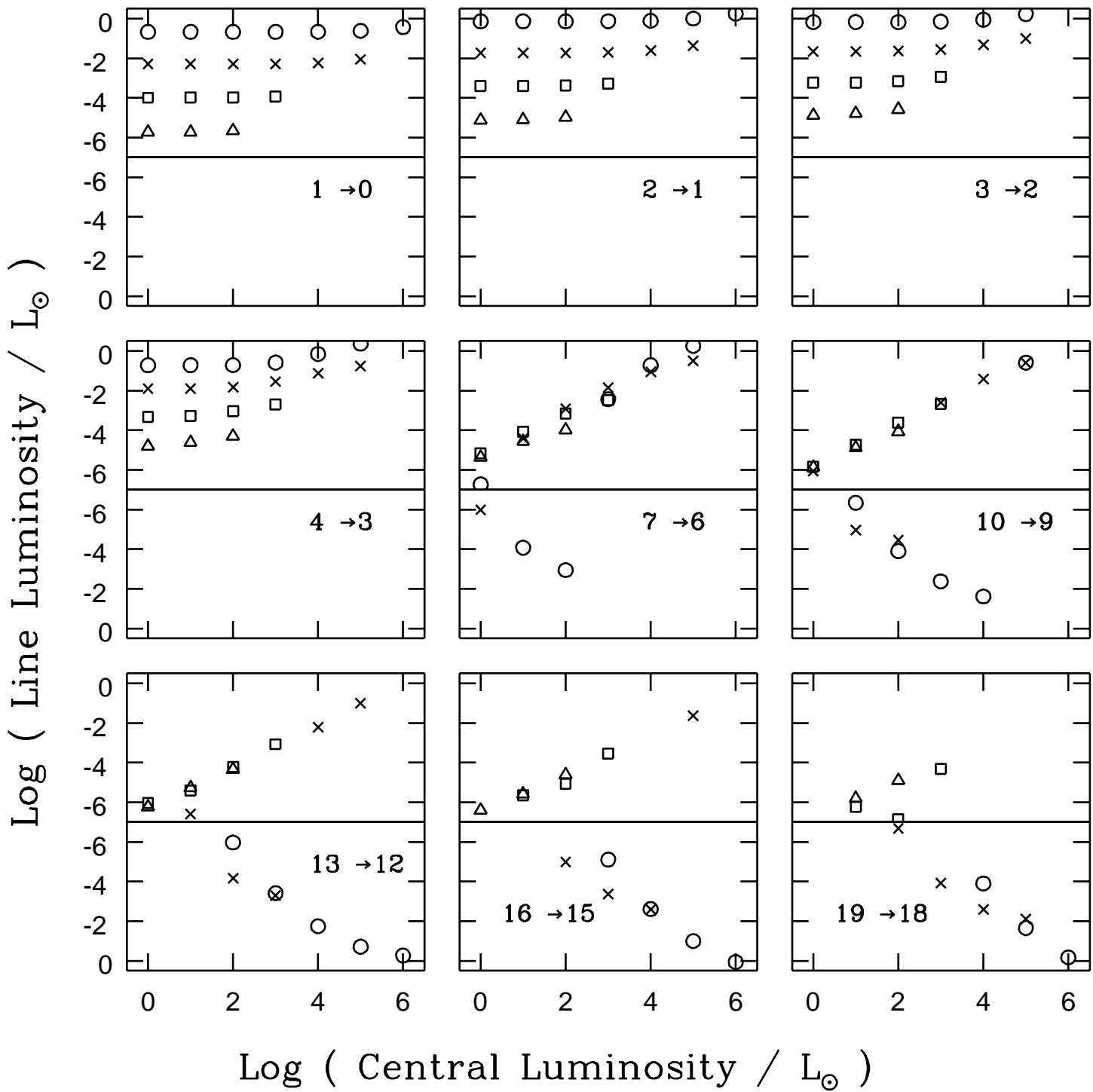
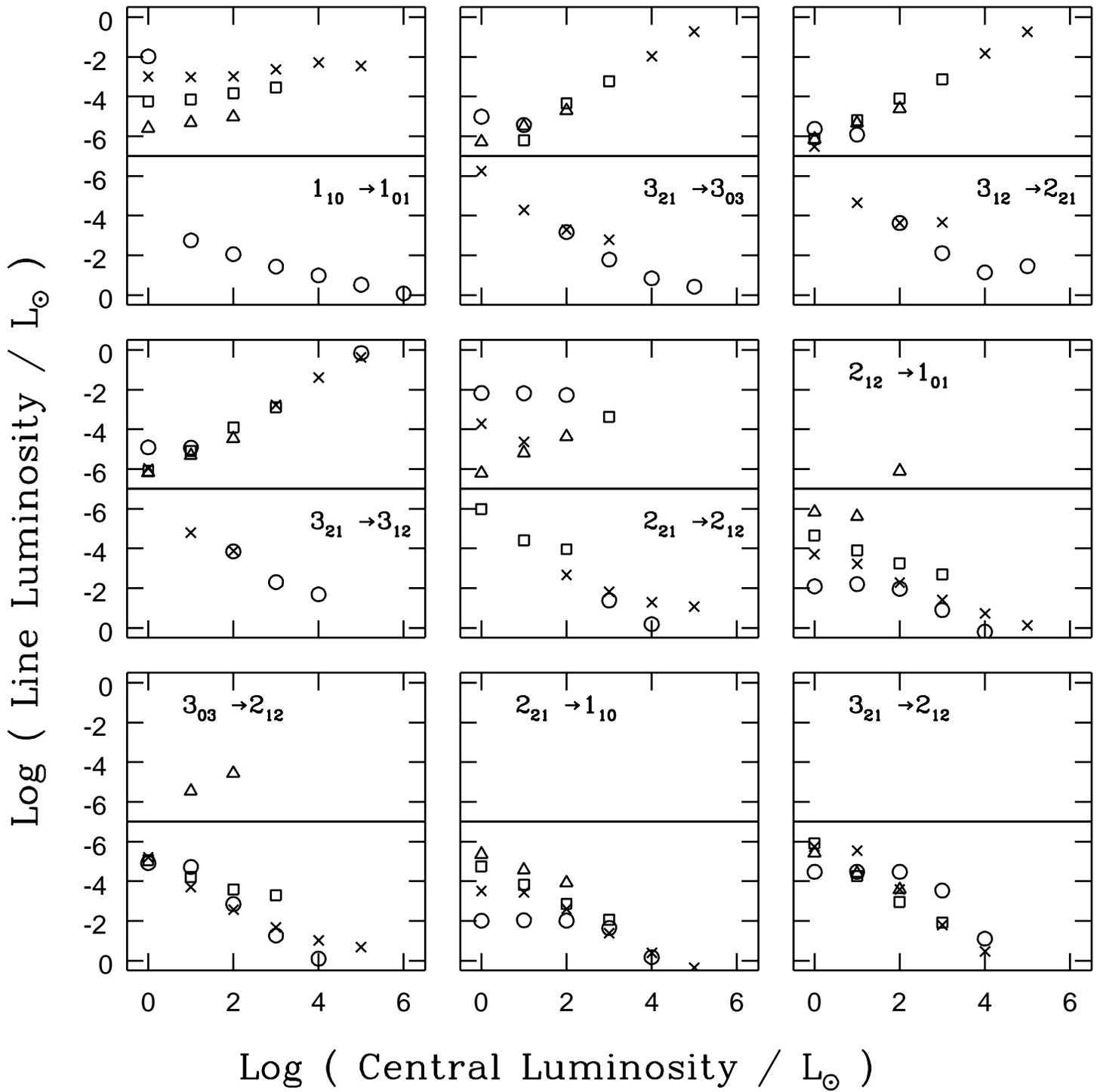
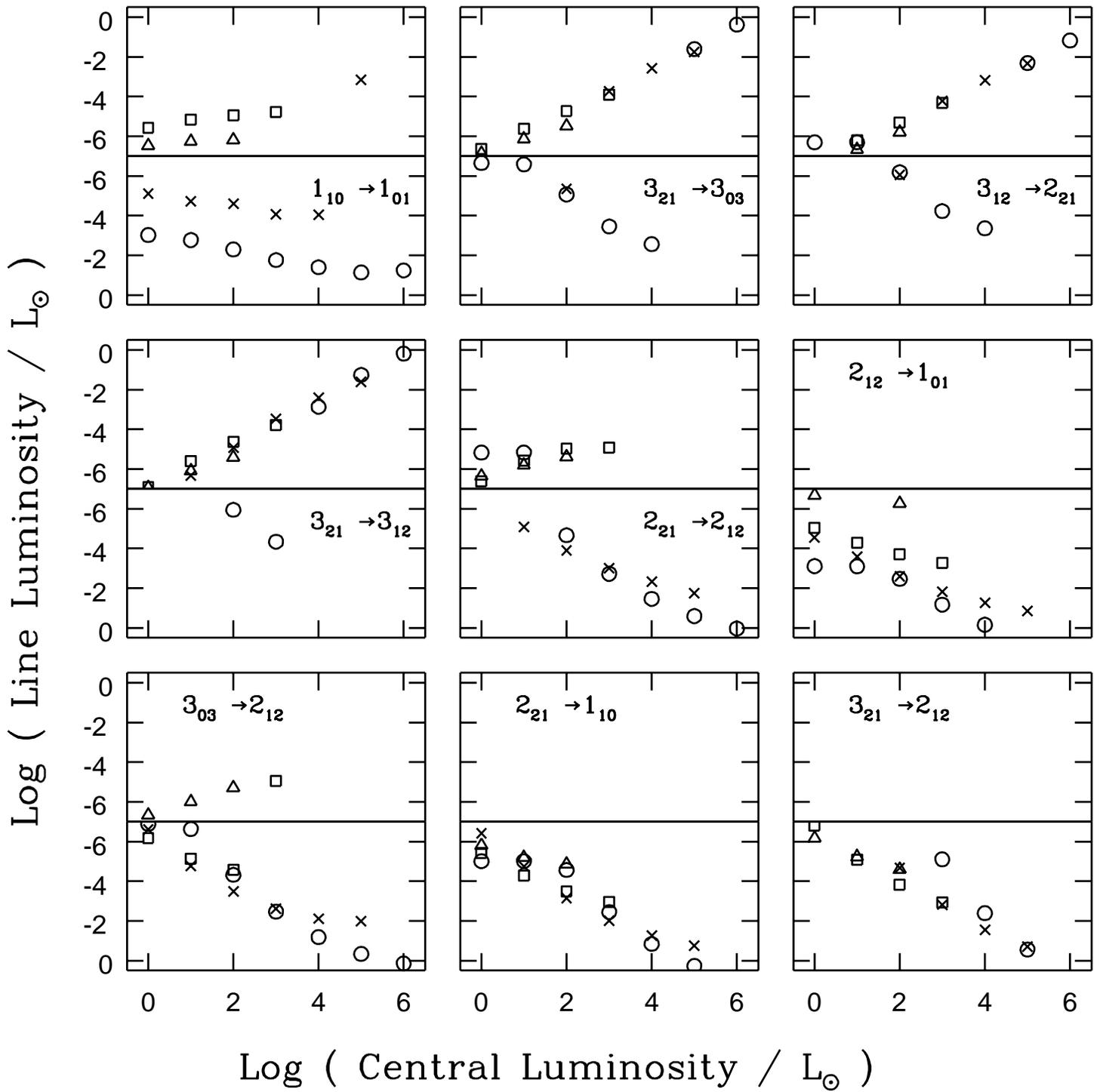
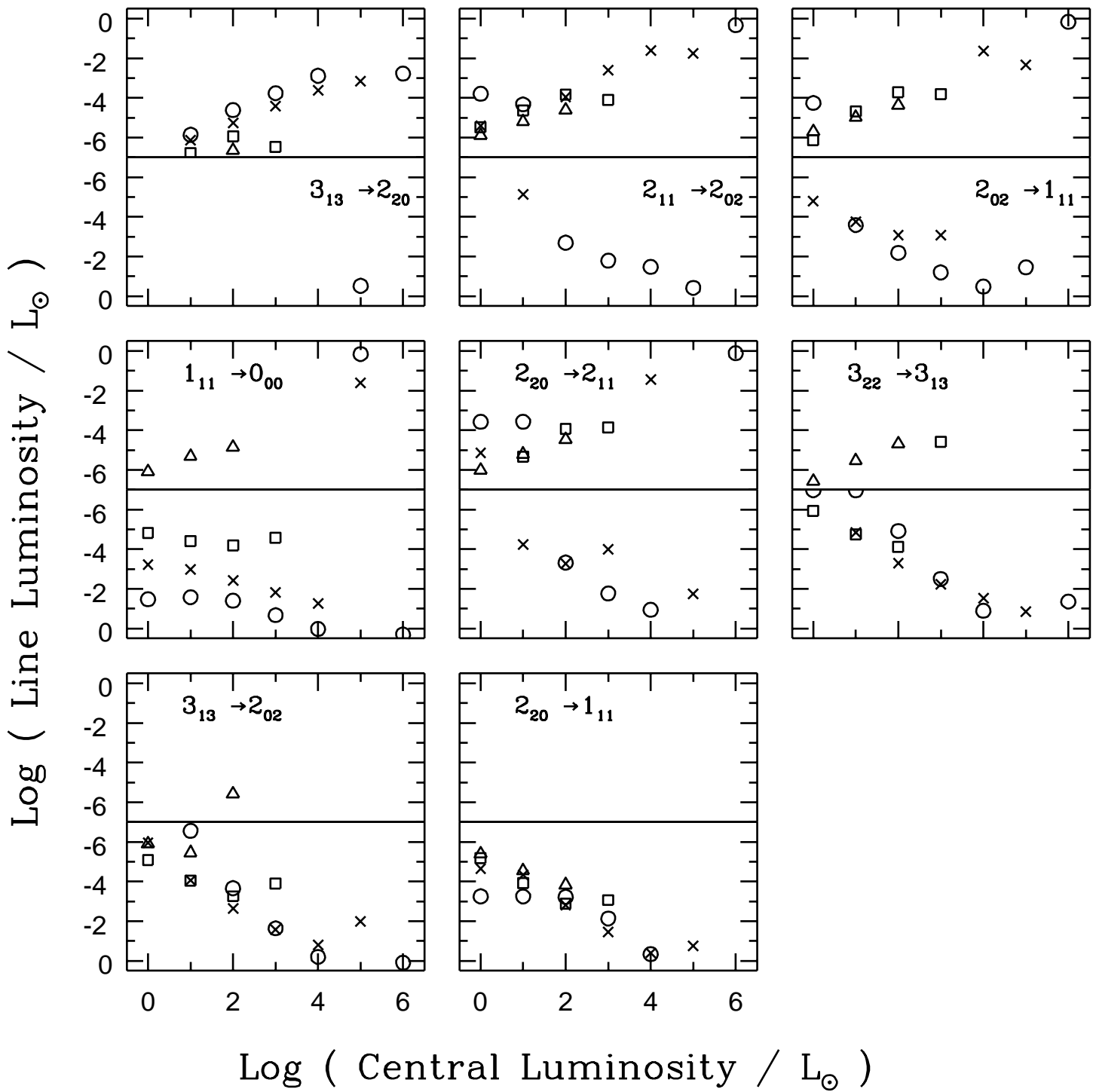
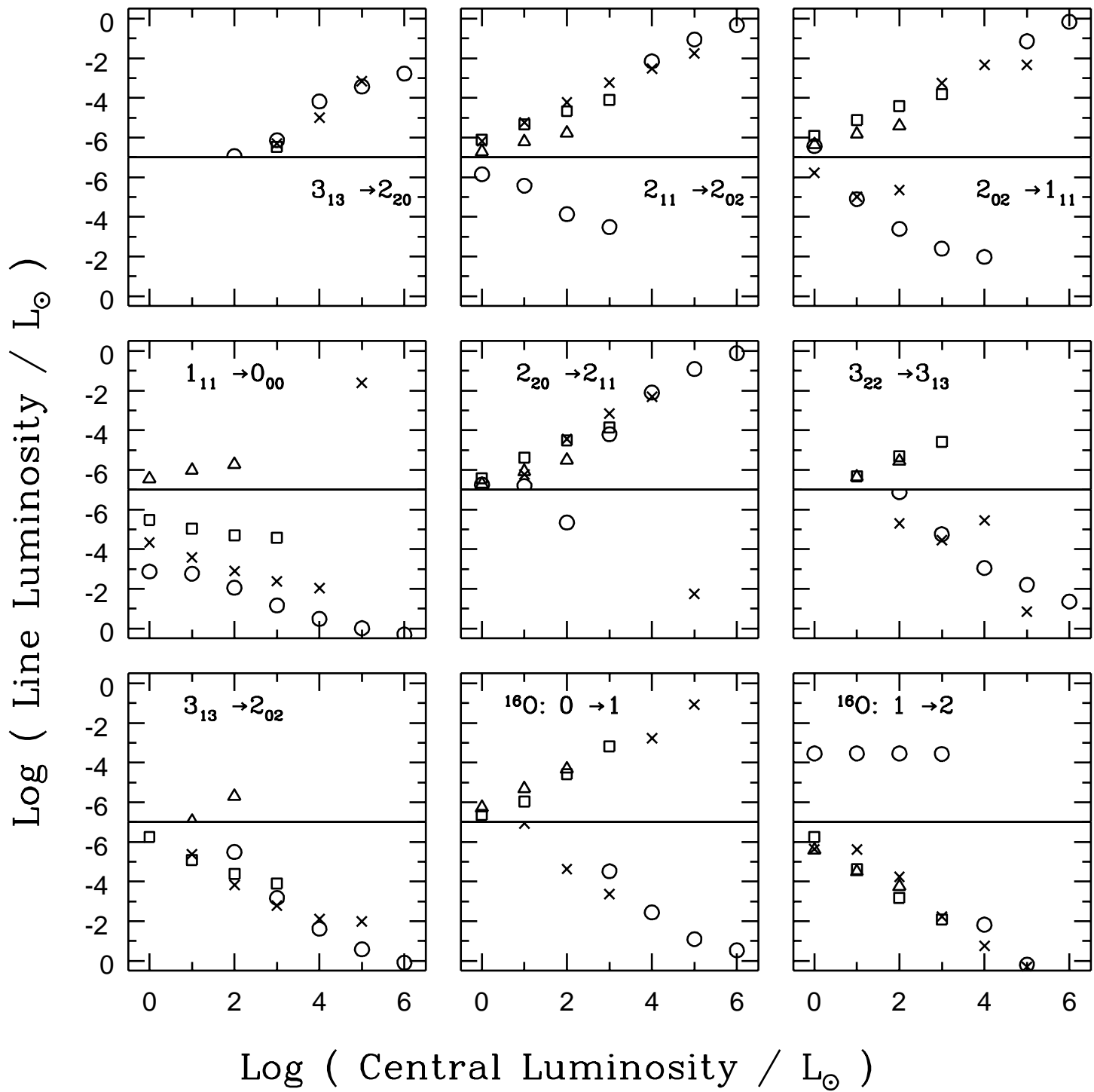
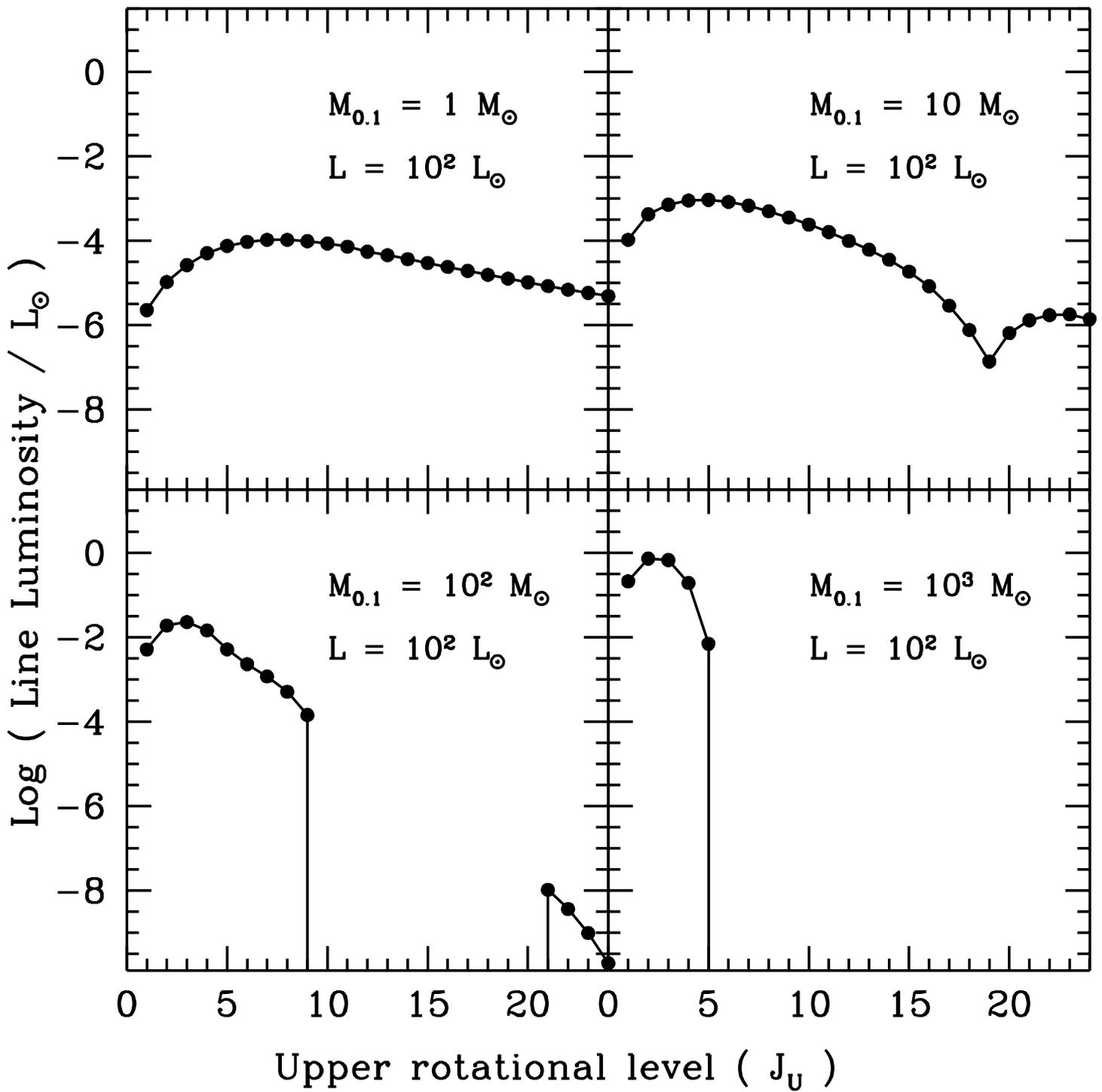
\end{document}